\documentclass[12pt]{article}

\usepackage[utf8]{inputenc}
\usepackage{amsmath}
\usepackage{amssymb}
\usepackage{amsfonts}
\usepackage[dvipdfmx]{graphicx}
\usepackage{color}
\usepackage{amsbsy}
\usepackage{url}
\usepackage{enumerate}
\usepackage[vcentermath,enableskew]{youngtab}
\usepackage{verbatim}
\usepackage[normalem]{ulem}
\usepackage{cancel}
\usepackage{bm,bbm}
\usepackage{slashed}
\usepackage{subcaption}
\usepackage{hyperref}
\usepackage{cancel}
\usepackage{cite}

\usepackage{setspace}

\makeatletter
\@addtoreset{equation}{section}
\renewcommand{\theequation}{\thesection.\arabic{equation}}
\makeatother

\newcommand{\diag}{\mathrm{diag}}

\newcommand{\acal}{\mathcal{A}}
\newcommand{\dcal}{\mathcal{D}}
\newcommand{\tr}{\mathrm{Tr}}
\newcommand{\lcal}{\mathcal{L}}
\newcommand{\Acal}{\mathcal{A}}
\newcommand{\xcal}{\mathcal{X}}

\newcommand{\fcal}{\mathcal{F}}

\newcommand{\refer}[1]{(\ref{#1})}
\newcommand{\diff}{\mathrm{d}}

\newcommand{\Tr}{\mathrm{Tr}}

\newcommand{\threetwo}{$\bf{3\mbox{-}2}$ }

\def\be{\begin{eqnarray}}
\def\ee{\end{eqnarray}}
\def\p{\partial}

\begin{document}

\begin{titlepage}

\begin{center}
 
{\Large \bf Five benefits of grand unified  
$SU(5)$ brane world scenario}
\vskip 1.0cm
\normalsize

 {\bf Masato Arai$^{a}$\footnote{arai(at)sci.kj.yamagata-u.ac.jp},
Filip Blaschke$^{bc}$\footnote{filip.blaschke(at)physics.slu.cz},
Minoru Eto$^{ade}$\footnote{meto(at)sci.kj.yamagata-u.ac.jp}, 
Masaki Kawaguchi\footnote{ddwbb.daigaku@gmail.com},
\\
}
\vskip 0.2cm

  { \it 
 $^a$Faculty of Science, Yamagata University, 
Kojirakawa-machi 1-4-12, Yamagata,
Yamagata 990-8560, Japan\\
$^b$Research Centre for Theoretical Physics and Astrophysics, Institute of Physics, 
Silesian University in Opava, Bezru\v{c}ovo n\'am. 1150/13, 746~01 Opava, Czech Republic. \\
$^c$Institute of Experimental and Applied Physics, Czech Technical University in Prague, Husova~240/5, 110~00 Prague~1, Czech Republic\\
$^d$Department of Physics, and Research and 
Education Center for Natural Sciences, 
Keio University, 4-1-1 Hiyoshi, Yokohama, Kanagawa 223-8521, Japan\\
$^e$International Institute for Sustainability with Knotted Chiral Meta
Matter(WPI-SKCM$^2$), Hiroshima University, 1-3-2 Kagamiyama,
Higashi-Hiroshima, Hiroshima 739-8511, Japan
}

\vskip 0.5 cm

\begin{abstract}
We construct an $SU(5)$ Grand Unified Theory on domain walls in the five-dimensional space-time.
In this setup, we introduce an adjoint scalar field and a singlet that 
 together form a set of five domain-wall solutions, realizing a dynamical brane-world.
The same scalar fields also localize chiral fermion zero modes around the walls via the Jackiw-Rebbi mechanism, 
 break $SU(5)$ down to the Standard Model gauge group via geometric Higgs mechanism and simultaneously trap gauge fields 
 through a field-dependent gauge kinetic term.
 Furthermore, they enable localization of the Higgs field, providing a novel solution to the doublet-triplet 
 splitting problem.
 As a result, all essential ingredients of the model are realized by a single adjoint scalar field and a singlet, 
 making the construction very economical.
We propose two realizations of the Higgs sector, derive the four-dimensional effective theory,
 and demonstrate that the Standard Model Yukawa couplings at the weak scale can be 
 reproduced from the five-dimensional Yukawa couplings by the renormalization group analysis with 
 a suitable choice of parameters.
 
 
\end{abstract}
\end{center}
\end{titlepage}

\clearpage
\thispagestyle{empty}
\vspace*{0.4\textheight}
\begin{center}
{\large \it Dedicated to the memory of Professor Norisuke Sakai.}
\end{center}
\clearpage

%
%
\section{Introduction}
The hypothesis of extra dimensions has been extensively explored in high-energy physics and cosmology.
Within this context, the brane-world framework provides a compelling paradigm in which our observable universe is 
 realized as a three-dimensional hypersurface (3-brane) embedded in a higher-dimensional space-time.
Prominent realizations of this idea include the model of Arkani-Hamed, Dimopoulos, and Dvali \cite{ArkaniHamed:1998rs},
 which introduces large compact extra dimensions to lower the fundamental Planck scale; 
its embedding into the superstring theory is studied in \cite{Antoniadis:1998ig}. 
The Randall-Sundrum model \cite{Randall:1999ee, Randall:1999vf} employs warped extra dimensions to generate 
 a hierarchy with an exponential factor.
Both scenarios illustrate how higher-dimensional geometries can offer novel mechanisms to address long-standing 
 puzzles of the Standard Model (SM), most notably the gauge hierarchy problem.

However, this scenario suffers from fundamental problems: the existence of branes and the localization of SM 
 fields are simply assumed.
These problems can be resolved by introducing topological solitons.
The dynamical origin of branes in extra dimensions arises from spontaneous symmetry breaking, which generates a topologically 
 stable soliton that hosts our four-dimensional universe.
This topological structure not only guarantees the stability of the brane but also naturally explains the localization of 
 chiral matter fields \cite{Rubakov:1983bb, Jackiw:1975fn}.
Remarkably, gravitons can likewise be trapped \cite{Cvetic:1992bf,DeWolfe:1999cp,
Csaki:2000fc,Eto:2002ns,Eto:2003ut, Eto:2003bn}, ensuring that both matter and gravity are confined to 
 the same four-dimensional surface.
Thus, topological solitons provide a compelling and elegant mechanism that bridges the gap between higher-dimensional 
 theories and the observed four-dimensional world.

Despite this success, the localization of massless gauge bosons remains a highly nontrivial problem. 
When the gauge symmetry $H$ is spontaneously broken in the bulk (the Higgs phase) and restored only in the vicinity of 
 a topological soliton, one might anticipate the emergence of massless gauge bosons localized within the soliton's core.
However, this expectation is not realized in general: massless modes acquire masses proportional to the inverse 
 width of the soliton~\cite{Dvali:1996xe}.
The reason is that the bulk region, apart from the soliton core, behaves as a non-Abelian superconductor.
Although the gauge symmetry is restored inside the soliton, any electric flux generated by a charged probe placed within 
 it is absorbed into the surrounding superconducting medium.
Consequently, gauge fields can propagate only over distances comparable to the soliton width, implying that they are 
 massive~\cite{Antoniadis:1998ig, Dvali:1996xe}.
This reasoning is physically well-motivated, and thus, the localization of massless gauge bosons on a soliton appears, in general, to be a highly nontrivial challenge.

A crucial insight to overcome this difficulty was introduced in Ref.~\cite{Dvali:1996xe}, where a mechanism based on 
 electromagnetic duality was proposed.
  \footnote{Following \cite{Dvali:1996xe}, numerous studies on the localization of massless gauge bosons have been 
  carried out \cite{Dvali:2000rx, Kehagias:2000au, Dubovsky:2001pe, 
  Ghoroku:2001zu,Akhmedov:2001ny, Kogan:2001wp, Abe:2002rj, 
  Laine:2002rh, Maru:2003mx, Batell:2006dp, Guerrero:2009ac, 
  Cruz:2010zz, Chumbes:2011zt, Germani:2011cv, Delsate:2011aa, 
  Cruz:2012kd, Herrera-Aguilar:2014oua, Zhao:2014gka, Vaquera-Araujo:2014tia,
  Alencar:2014moa,Alencar:2015awa,Alencar:2015oka,Alencar:2015rtc,Alencar:2017dqb,Pantoja:2018edw, Fu:2026zyf}.
 Each offers certain advantages but also facing specific limitations, and a comprehensive understanding is still lacking.}
If a probe magnetic charge is placed inside the soliton embedded in a superconducting bulk, the magnetic flux is expelled 
 from the bulk by the Meissner effect.
Owing to flux conservation, the field lines are then confined to extend along the soliton, effectively to 
 infinity~\cite{Antoniadis:1998ig, Dvali:1996xe}.
In Ref.~\cite{Dvali:1996xe}, this situation was reformulated in a dual picture: the Higgs phase in the bulk is replaced by a 
 confining phase, realized by the condensation of magnetic charges.
Under this assumption, a massless gauge field can be localized inside the soliton when 
 the  gauge symmetry $H$ remains unbroken within the soliton but is embedded into a larger non-Abelian group $G$
 in the confining bulk.
This localization mechanism, known as the Dvali-Shifman (DS) mechanism, was rigorously demonstrated in four-dimensional 
 super Yang-Mills theory~\cite{Dvali:1996xe}.
Its extension to higher dimensions, however, remains unproven, as the nature of confinement in dimensions beyond four 
 is not yet fully understood.
Following Ref.~\cite{Dvali:1996xe}, fat-brane scenarios based on this idea have been studied \cite{Davies:2007xr, Davidson:2007cf, Thompson:2009uk, Callen:2010mx}.
However, most of these works assume that the DS mechanism is valid.

On the other hand, a simpler mechanism for gauge field localization was proposed by Ohta and Sakai \cite{Ohta:2010fu}.
In this mechanism, a field-dependent gauge kinetic term (field-dependent permeability) is introduced:
\begin{eqnarray}
 {\cal L}_{\rm gauge}=-\beta(T)^2F_{MN}F^{MN}\,,
 \label{eq:OS_mechanism}
\end{eqnarray}
where $M, N$ denote indices of the full space-time and $T$ is a scalar field.
The field strength is given as $\fcal_{MN} = \partial_{M} \acal_{N} - \partial_{N} \acal_{M} + i [ \acal_{M} , \acal_{N}]$. 
This represents a semiclassical realization of the confining phase 
 \cite{ArkaniHamed:1998rs,Dvali:1996xe,Kogut:1974sn,Fukuda:1977wj, Luty:2002hj, Fukuda:2009zz, Fukuda:2008mz}, 
 rather than the Higgs phase that exists outside the solitons.
When $\beta(T)^2$ is square-integrable over the entire extra-dimensional space, the massless four-dimensional gauge fields 
 become localized on the topological soliton.
Because square-integrability is independent of the specific details of a model, the resulting localization mechanism is quite robust.
We have confirmed this robustness in a number of explicit five- and six-dimensional models  \cite{Arai:2012cx,Arai:2013mwa,Arai:2017lfv,Arai:2017ntb,Arai:2018uoy, Arai:2021wul}
 and have also provided a formal proof valid in arbitrary spacetime dimensions \cite{Arai:2018rwf, Eto:2019weg}.
The same localization mechanism is used to construct the five-dimensional brane world models in 
 \cite{Okada:2017omx, Okada:2018von, Okada:2019fgm, Das:2025wow}.

In \cite{Arai:2017lfv}, we have constructed an $SU(5)$ Grand Unified Theory (GUT) in five dimensions.
In this model, two types of scalar fields $S$ and $T$ are introduced, where the $T$ sector consists of an adjoint scalar field $\hat{T}$ 
 and a singlet scalar field $T^0$. 
The scalar field $S$ is responsible for localizing gauge fields 
and $T$ is for localizing matter fields.
More specifically, 
$T$ sector plays three important roles: 1) it generates the domain-wall configuration, 
 2) it consequently traps chiral fermion zero modes around the wall,
 and  3) it breaks $SU(5)$ down to the SM gauge group.
It was further shown that proton decay is suppressed due to the reduced overlap of the 
 zero mode functions
of the fermions and gauge fields.
However, the $SU(5)$ model proposed in \cite{Arai:2017lfv} remains phenomenologically incomplete, since neither the Higgs sector nor the Yukawa couplings were incorporated.
As a result, it was not possible to address electroweak symmetry breaking or to derive the mass spectrum of the matter fields.
Moreover, the model could not deal with the doublet-triplet splitting problem in the $SU(5)$ GUT.

In this paper, we construct a five-dimensional $SU(5)$ brane-world model based on domain-wall solutions with a Higgs 
 sector and Yukawa couplings.
In contrast to the previous model \cite{Arai:2017lfv}, 
the present framework employs 
only the $T$ sector, consisting of the adjoint scalar field $\hat T$ and the singlet scalar field $T^0$, without introducing the additional scalar field $S$.
The $T$ sector is responsible 
 not only for the three tasks 1) - 3) explained above but also for 4) the localization of 
 gauge fields which was 
 achieved
 by the $S$ field in \cite{Arai:2017lfv}, making the construction more economical.
 Furthermore, we will show that the $T$ field can also contribute to 5) localization of the SM Higgs field on the walls avoiding the doublet-triplet splitting problem.
Theories with extra dimensions provide an attractive framework to solve the doublet-triplet splitting problem 
 \cite{Kawamura:2000ev, Kakizaki:2001en, Hebecker:2001wq, Maru:2001ch, Haba:2003jt, Haba:2002if, deAnda:2023spb}.  
Within the higher-dimensional context, we propose two distinct realizations of the Higgs sector that successfully resolve the 
 splitting problem.
We further derive the four-dimensional effective theory and show that the
 SM Yukawa couplings can be reproduced from the five-dimensional
 Yukawa interactions through renormalization group evolution of the SM.

An additional advantage of the present setup concerns the fermion mass structure. 
In minimal four-dimensional $SU(5)$ GUTs, the unified Yukawa structure
 generically predicts the relation $Y_d = Y_e^{T}$ at the unification scale, 
 where $Y_d$ and $Y_e$ denote the Yukawa matrices for down-type quarks and
charged leptons, respectively.
This relation leads to phenomenologically unrealistic fermion mass relations.
In our framework this problem is avoided without 
extending the Higgs sector (for instance by introducing additional Higgs representations 
 such as a $\mathbf{45}$ as in the Georgi-Jarlskog mechanism \cite{Georgi:1979df}) or invoking 
 higher-dimensional operators that modify the Yukawa structure \cite{Ellis:1979fg}.
Although the underlying five-dimensional Yukawa couplings respect $SU(5)$
 unification, the effective four-dimensional Yukawa matrices arise from overlap
 integrals of localized fermion zero modes in the extra dimension.
These generation-dependent overlaps naturally relax the usual $SU(5)$ mass
 relations, allowing a realistic pattern of fermion masses and mixings.

The organization of this paper is as follows.
In Sec.~\ref{sec:model-tot}, we provide an overview of the model. Sec.~\ref{sec:model} is devoted to the discussion 
 of the domain-wall sector.
In Sec.~\ref{sec:Higgs}, we describe the Higgs sector, with particular focus on the two models that provide solutions to 
 the doublet-triplet splitting problem. Specifically, 
Sec.~\ref{sec:FerYu} presents the remaining sectors and the derivation of the four-dimensional effective theory of the model;
SubSec.~\ref{sec:gauge_eff} deals with the gauge sector and its four-dimensional effective theory,
SubSec.~\ref{sec:Higgs_eff} derives the four-dimensional effective theories of the two Higgs-sector models, 
SubSec.~\ref{sec:fer_eff} discusses the fermion sector and its effective theory, and
Sec.~\ref{sec:yukawa_eff} covers the Yukawa sector and the 
 corresponding four-dimensional effective theory.
In Sec.~\ref{sec:conc} we summarize our results.
In \ref{app:A}, we provide a derivation of the effective potential in the domain-wall sector. 
\ref{app:B} gives proofs of the decompositions used in Eq.~(\ref{eq:sum-A}) and Eq.~(\ref{eq:lag-x-phi-1}). 
\ref{app:C} summarizes the Standard Model renormalization group equations used in the analysis of the Yukawa couplings in 
 Sec.~\ref{sec:yukawa_eff}.

%
%
\section{Model}\label{sec:model-tot}
We consider a five-dimensional $SU(5)$ GUT with a Higgs ${\cal H}$ in a fundamental representation, 
  fermionic matters $\Psi_{\bar{5},A}$ in an anti-fundamental representation and 
  $\Psi_{10,A}$ in an antisymmetric representation, where the index $A=1,2,3$ denotes the generation. 
In addition, we introduce two scalar fields $ \hat{T} $ in an adjoint representation and a singlet $T^{0} $.
They transform under the gauge transformation with $U\in SU(5)$ as
\begin{eqnarray}
&&{\cal H}\rightarrow U{\cal H}\,, \\
&&\Psi_{\bar{5}}\rightarrow U^*\Psi_{\bar{5}}\,, \\
&&\Psi_{10}\rightarrow U\Psi_{10}U^t\,, \\
&&\hat{T}\rightarrow U\hat{T}U^\dagger\,.
\end{eqnarray}  
We decompose the full Lagrangian into the following parts:
\begin{eqnarray}
 {\cal L}={\cal L}_{\rm T}+{\cal L}_{\rm H}+{\cal L}_{\rm gauge}+{\cal L}_{\rm f}+{\cal L}_{\rm yukawa}\,. \label{total}
\end{eqnarray}
We briefly explain each part in the following.

The first part, ${\cal L}_T$, serves a role of providing domain wall background on which the SM is
 realized in the low energy limit.
This background allows for several configurations of domain walls and of spontaneous gauge symmetry breaking patterns.
We will show that the symmetry breaking from $SU(5)$ to $SU(3)\times SU(2)\times U(1)$ is realized as the most energetically favored
 for robust choices of parameters in the model.
We will provide an explicit form of ${\cal L}_T$ and a detailed analysis to obtain the domain walls in Sec.~\ref{sec:model}.
The effective Lagrangian of this sector will be discussed together with that of the gauge sector in Sec.~\ref{sec:gauge_eff}.

The second part, ${\cal L}_{\rm H}$, is the Lagrangian for the Higgs fields, which is one of the main focuses of this paper. 
This part provides a mechanism to localize the doublet part of ${\cal H}$, which becomes the SM Higgs field,
 while the triplet part remains delocalized or only appears in the low-energy effective theory through very heavy modes.
This is a solution to the doublet-triplet splitting problem.
We will propose two models to solve this problem in Secs.~\ref{sec:HSWOF} and \ref{sec:HSWF}.
We will also derive the effective Lagrangian of these two models in Sec.~\ref{sec:Higgs_eff}. 

The third part, ${\cal L}_{\rm gauge}$, describes the kinetic term of the $SU(5)$ gauge field 
 and  the  localization of SM gauge fields around a certain region of the fifth-dimension through the Ohta-Sakai mechanism \cite{Ohta:2010fu}.
In this Lagrangian, a non-trivial function of $\hat{T}$ and $T^0$ appears as a coefficient of the gauge kinetic term.
This means that the coefficient which is normally an inverse of a gauge coupling constant is 
 generalized to a positive semi-definite function of $\hat{T}$ and $T_0$.
This function plays an important role to localize gauge fields. 
We will describe ${\cal L}_{\rm gauge}$ in Sec. \ref{sec:gauge_eff}, where the effective Lagrangian of ${\cal L}_{\rm gauge}$ will be
derived as well.

The fourth part, ${\cal L}_{\rm f}$, includes the kinetic terms of the fermion fields and their couplings to the $\hat T$ and $T_0$ fields, by which the fermion
 can localize around the domain wall. 
The fifth part, ${\cal L}_{\rm yukawa}$, contains the 5-dimensional Yukawa terms that give the masses to the fermions 
 after the electroweak symmetry breaking.
These two sectors will be discussed, including the derivation of their effective Lagrangians, in Secs.~\ref{sec:fer_eff} and 
 \ref{sec:yukawa_eff}.
In Sec.~\ref{sec:yukawa_eff}, we perform the renormalization group analysis of the Yukawa couplings and show 
 that the Yukawa couplings of the SM at $Z$-boson scale are realized by suitable choice of the Yukawa couplings 
 and parameters in the five-dimensional model.

%
%
\section{Domain wall sector}\label{sec:model}
\subsection{Domain wall background}
For convenience, let us combine the $SU(5)$ adjoint field $\hat T$ and the scalar field $T_0$ into a single entity $T$:
\begin{equation}
T \equiv \hat T +\frac{T_0}{5}\mathbf{1}_5\,.
\end{equation}
We then organize the $\mathcal{L}_{\rm T}$ as follows\footnote{We take the metric to be $\eta_{MN}={\rm diag}(1,-1,-1,-1,-1)$, where $M, N=0, \cdots 3, 4$ stand for the five-dimensional 
 space-time indices.}
\begin{eqnarray}
 \lcal_{\rm T} = \tr \left [\dcal_{M} T \dcal^{M} T  \right ] - V_0 - V_1\,, \label{eq:32model}
\end{eqnarray}
where the covariant derivative is given by $\dcal_M T = \p_M T + i [{\cal A}_M,T]$ with ${\cal A}_M$ being an $SU(5)$ gauge field.
Note that the kinetic term is not canonically normalized for $T_0$ field, i.e. $ \tr \left [\dcal_{M} T \dcal^{M} T  \right ] =  \tr \left [\dcal_{M} \hat T \dcal^{M} \hat T  \right ]+ \tfrac{1}{5}\partial_M T_0 \partial^M T_0$.

The first part of the potential $V_0$ is chosen in such a way as to facilitate domain walls (and anti-walls) in the background and it is assumed that it is dominant over the second part. The role of $V_1$ is to merely modify the shape and dynamics of the (anti-)walls slightly and mostly in a generic manner.
In particular, let us choose
\begin{equation}
V_0 =  \lambda\, \tr \bigl(v^2\mathbf{1}_5-T^2\bigr)^2\,.
\end{equation}

If we set $V_1 = 0$ for the moment, it is easy to see how $V_0$ gives rise to domain (anti-)walls. First, we observe that there are $2^5$ discrete vacua described by 
\begin{equation}\label{eq:vacs}
T_{\rm vac} \equiv v\, U \mbox{diag}\bigl(s_1, s_2, \ldots, s_5\bigr) U^\dagger\,,
\end{equation} 
where $U$ is an arbitrary $SU(5)$  matrix and $s_i \in \{-1,1\}$ are a quintuplet of signs, while $v$ is the vacuum expectation value.

For further purpose, let us organize the canonical vacua, given by Eq.~\refer{eq:vacs} with fixed $U = \mathbf{1}_5$, into three 
 sets $\langle 0 \rangle $, $\langle 1 \rangle $ and $\langle 2 \rangle $  according to how they break $SU(5)$ gauge symmetry:
 {\small \begin{align}
\langle 0 \rangle & \equiv T_{\rm vac}/v \in \left\{ \mathbf{1}_5, -\mathbf{1}_5\right\} & & \mbox{$SU(5)$ preserving} \\
\langle 1 \rangle & \equiv T_{\rm vac}/v \in \left\{ \begin{pmatrix}-1 & \\ & \mathbf{1}_4\end{pmatrix},  \begin{pmatrix}1 & \\ & -\mathbf{1}_4\end{pmatrix}, \mbox{$+$ permutations}\right\} & & SU(5)\to U(1)\times SU(4) \\
\langle 2 \rangle & \equiv T_{\rm vac}/v \in \left\{ \begin{pmatrix}-\mathbf{1}_2 & \\ & \mathbf{1}_3\end{pmatrix},  \begin{pmatrix}\mathbf{1}_2 & \\ & -\mathbf{1}_3\end{pmatrix}, \mbox{$+$ permutations}\right\} & & SU(5)\to U(1)\times SU(2)\times SU(3) 
\end{align}}
While there are only two elements in the set $\langle 0 \rangle$, there are 10 in $\langle 1 \rangle$ and 20 in $\langle 2 \rangle$, totalling  $2^5 = 32$ elements. 

Next let us consider the domain walls. Since all the vacua mentioned above are discrete, stable domain walls exist.
We choose the $SU(5)$ unbroken vacua $\langle 0 \rangle$ and take the following boundary conditions
\begin{align}
T \xrightarrow[y \rightarrow -\infty]{} - v {\bf 1}_5  , \quad T \xrightarrow[y \rightarrow \infty]{} v {\bf 1}_5. \label{BC of T}
\end{align}
A general static solution with these boundary conditions can be obtained in a close form as
\begin{equation}\label{eq:dsols}
T_{\rm walls} = v\, \mbox{diag}\,\bigl(\tanh\bigl(\Omega y-R_i\bigr)\bigr)\,, \hspace{5mm} i =1,\ldots ,5\,,
\end{equation}
where $\Omega \equiv \sqrt{\lambda}v$.
This solution describes a configuration of five domain walls at (dimensionless) positions $R_i$ along the the extra-dimensional coordinate $y \equiv x^4$. 
 
\begin{figure}[t]
\begin{center}
\includegraphics[width=0.5\columnwidth]{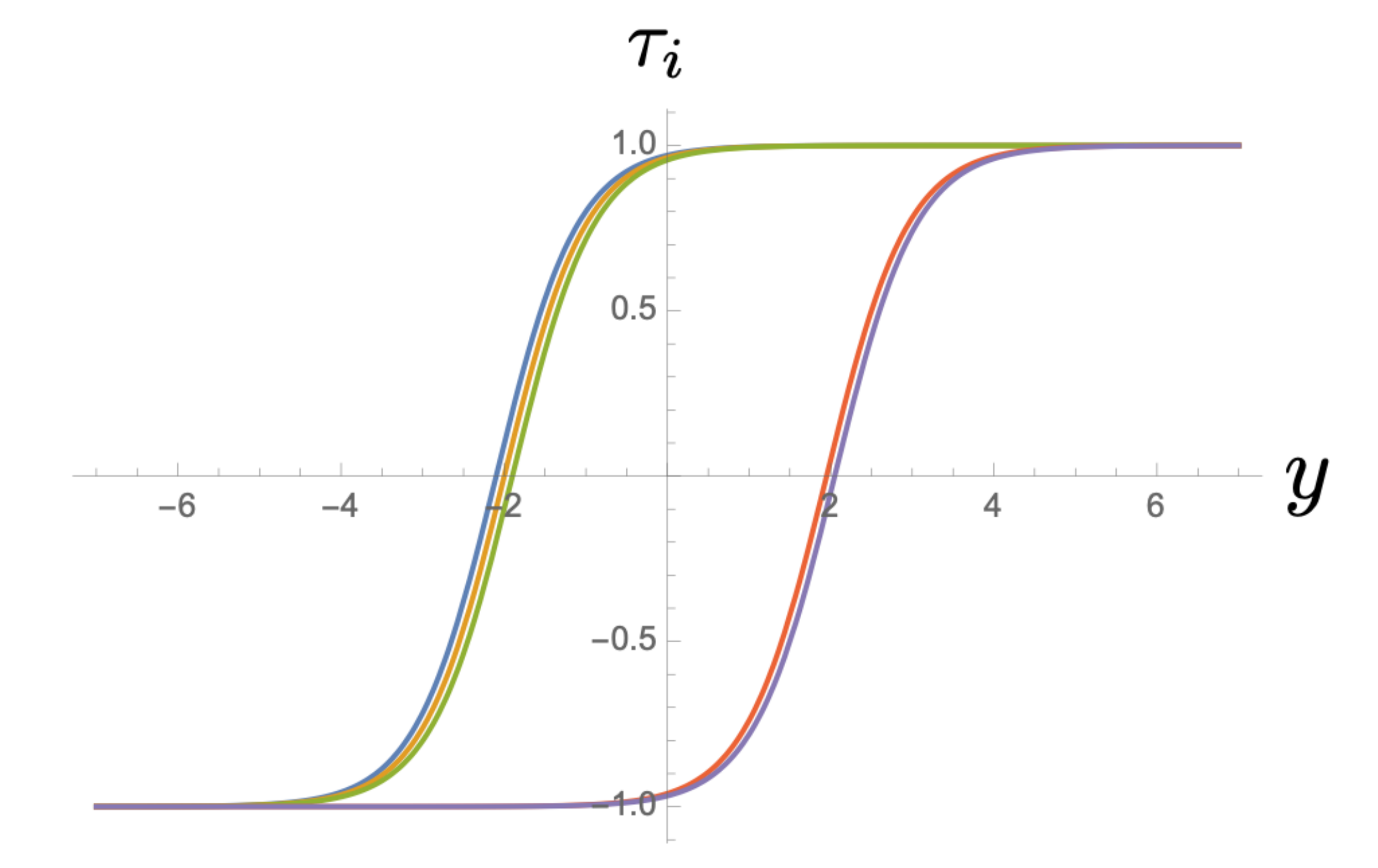}
\caption{\small Illustration of the  ${\bf 3}$-${\bf 2}$ split configuration of domain walls. Here, $v=\Omega=1$ and the position moduli are given as $R_1=R_2=R_3=-2, R_4=R_5=2$.}
\label{domain2}
\end{center}
\end{figure}

Depending on the number of coincident walls and the ordering of their positions on the $y$-axis, there are ten patterns of possible configurations.
A phenomenologically interesting case is when three coincident walls and two coincident domain walls are 
 placed at different positions, namely, $R_1=R_2=R_3<R_4=R_5$.
We write this configuration as ${\bf 3}$-${\bf 2}$, where the numerals denote a number of coincident walls and the hyphen 
 means the space separation.
This configuration is depicted in Fig. \ref{domain2}. 
We can think about this configuration as consisting of regions of two asymptotic outer vacua and one approximate inner vacuum, in which different symmetry breaking patters are realized. 
The $SU(5)$ is unbroken far away from the domain walls, but it is spontaneously broken to $SU(3) \times SU(2) \times U(1)$
 in the middle region, where $T\approx v\,\diag(1,1,1,-1,-1)$.
There are, of course, many other configurations such as ${\bf 4}$-${\bf 1}$ and ${\bf 2}$-${\bf 1}$-${\bf 2}$, etc.
As we will see in the next subsection, the ${\bf 3}$-${\bf 2}$ configuration is energetically favored compared to any other as $V_1$ is switched. Furthermore, we will show that this remains true in a large region of the parameter space and hence represents a generic case.

%
%
\subsection{Moduli stabilization}
\label{sec:moduli_stabilization}

Let us now turn our attention to the question of how the choice of $V_1$ affects this picture. 
We consider the following potential.
\begin{eqnarray}
V_1 = \lambda (-\mu^2 v^2 \Tr\bigl(\hat T^2)+\alpha \Tr\bigl(\hat T^4\bigr))\,.
\end{eqnarray}
In fact, it turns out that we can be quite general in choosing the extra pieces, provided that they are small compared with $V_0$ and hence can be treated perturbatively. Before we discuss generic case, however, let us illustrate the situation with specific examples.

This potential (here $\mu$ and $\alpha$ are dimensionless couplings) can be characterized as an extra double-well potential for the adjoint part $\hat T $. First of all, it changes the structure of the vacua. Namely, the additional energy due to $V_1$ for the three sets of vacua defined in the previous subsection are given as
 \begin{align}
 V_1\bigl(\langle 0 \rangle\bigr) = &\  0\,,  \\
 \label{eq:vac1} V_1\bigl(\langle 1 \rangle\bigr) = &\  \frac{16 \lambda v^4}{125}\bigl(52\alpha -25\mu^2\bigr)\,,  \\
 \label{eq:vac2} V_1\bigl(\langle 2 \rangle\bigr) = &\  \frac{24 \lambda v^4}{125}\bigl(28\alpha -25\mu^2\bigr)\,.
 \end{align}   
As we see, which set corresponds to a minimum depends on the values of $\alpha$ and $\mu$. 
In the following, we consider $\alpha > 25 \mu^2/28$, which realizes a phenomenologically interesting case, namely 
that $SU(5)$-preserving vacua are the true minima of the full potential with the total energy $V_0\bigl(\langle 0 \rangle\bigr)+V_1\bigl(\langle 0 \rangle\bigr)  = 0$.
As there are only two vacua in the $\langle 0 \rangle$ set, all diagonal entries of the $T$ field must be either walls or anti-walls. Without the loss of generality, let us take $s_1 = \ldots = s_5 = 1$.
 
Treating $V_1$ as perturbation, we can ascertain  which configuration of static domain walls has the lowest energy by plugging (\ref{eq:dsols})
into $V_1$ and integrating the result over $y$. 
Furthermore, we obtain effective potential as a function of distances between walls $R_i - R_j$:
\begin{equation}
V_{\rm eff}(R_i-R_j) \equiv \int\limits_{-\infty}^{\infty}\diff y\, V_1\bigl(\hat T_{\rm walls}\bigr)\,.
\end{equation}
Notice that due to invariance of $V_1$ with respect to permutation of diagonal entries, we can assume $R_1 < R_2 < \ldots < R_5$ without loss of generality. 
The effective potential $V_{\rm eff}(R_i-R_j) $ can be obtained exactly, which we do in App.~\ref{app:A}. 

However, for our purposes, let us demonstrate our claim using a simple argument. Naively, we may be tempted to think that a fully coincident configuration of walls, i.e.
\begin{equation}
T_{\rm coin} = v \tanh(x) \mathbf{1}_5\,,
\end{equation}
is the minimum of the energy, since $V_1\bigl(\hat T_{\rm coin}\bigr) = 0$, 
 where $x\equiv \Omega y$.
Departing slightly from the coincident point, however, we can show that $T_{\rm coin}$ must be in fact a local maximum. Indeed, given that $\hat T$ would be small, we can neglect the quartic term in $V_1$ and easily integrate the quadratic term (see \ref{app:A}) which gives
\begin{equation}
-\lambda v^2 \mu^2 \int\limits_{-\infty}^{\infty}\diff y\,  \Tr\bigl(\hat T_{\rm wall}^2\bigr) =  \sqrt{\lambda}\, v^3 \mu^2\Bigl(8-\frac{4}{5}\sum\limits_{i>j}\frac{R_i-R_j}{\tanh\bigl(R_i-R_j\bigr)}\Bigr)\,.
\end{equation}
As this is negative  -- as evident from the fact that the right-hand side is maximal at the fully coincident limit, $R_i = R_j$, for which it equals zero -- the effective potential implies a \emph{repulsive} force between any pair of walls for small separations. Hence, we see that the coincident point is unstable. 

The other extreme -- when all walls are well separated -- has the opposite effect, namely  an \emph{attractive} force due to the fact that field values in the gaps will be  
 in the $\langle 1 \rangle$ set for the two outer gaps and in the $\langle 2 \rangle$ set for the two inner gaps.
These gaps roughly gives a constant positive energy per unit length, which follows from (\ref{eq:vac1}) and (\ref{eq:vac2}).
Thus, 
these gaps 
would tend to collapse for large separations, but as we have seen, the fully coincident configuration is disfavored too. 
Hence, the true minimum should be a partially coincident configuration. 
Furthermore, the field value in the gap should be from the $\langle 2\rangle$ set,  since 
it is energetically favorable over the set $\langle 1\rangle$.

\begin{figure}[ht]
 \begin{center}
 \includegraphics[width=0.7\textwidth]{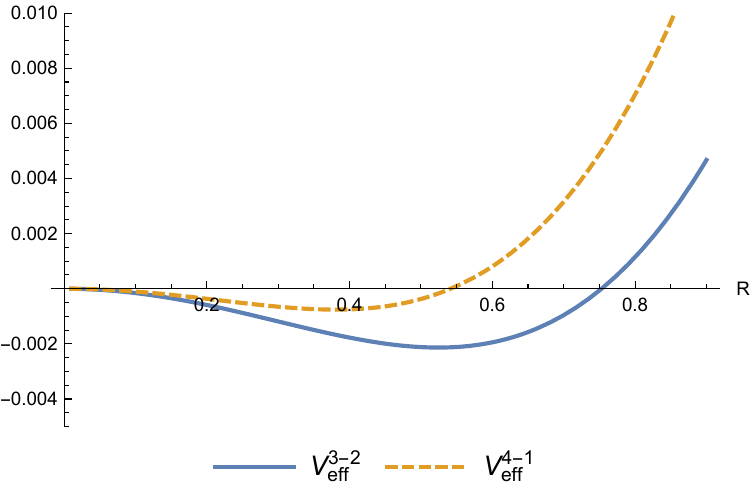}
 \caption{\small Comparison between effective potentails $V_{\rm eff}^{\bf{3}\mbox{-}\bf{2}}$ and $V_{\rm eff}^{\bf{4}\mbox{-}\bf{1}}$ showing that \threetwo splitting background is energetically preferable. Here, $ v= \lambda = 10 \alpha = 100 \mu^2 = 1$.}
 \label{fig:effpots}
 \end{center}
 \end{figure}

It is easy to establish that out of the two simplest possibilities -- the \threetwo splitting and the $\bf{4}\mbox{-}\bf{1}$ splitting configurations --
{\small \begin{equation}
T_{\bf{3}\mbox{-}\bf{2}}= v \,
\begin{pmatrix}
\tanh(x) \mathbf{1}_3 & \\ & \tanh(x-R)\mathbf{1}_2
\end{pmatrix}\,, \hspace{5mm}
T_{\bf{4}\mbox{-}\bf{1}}= v \, 
\begin{pmatrix}
\tanh(x) \mathbf{1}_4 & \\ & \tanh(x-R)
\end{pmatrix}\,, \label{eq:32and41}
\end{equation}}
the \threetwo splitting has the least energy. 
This can be checked by direct calculation of the effective potentials:
 \begin{align}
V_{\rm eff}^{\bf{3}\mbox{-}\bf{2}} \equiv  \int\limits_{-\infty}^{\infty}\diff y\, V_1\bigl(\hat T_{\bf{3}\mbox{-}\bf{2}}\bigr) = &
 \frac{112 v^3 \sqrt{\lambda} \alpha}{125}\biggl(-11+\frac{6R}{\tanh(R)}-\frac{15}{\sinh^2(R)}\Bigl(1-\frac{R}{\tanh(R)}\Bigr)\biggr)\nonumber \\ & + \frac{24 v^3 \sqrt{\lambda} \mu^2}{5}\Bigl(1-\frac{R}{\tanh(R)}\Bigr)\,, \\
 V_{\rm eff}^{\bf{4}\mbox{-}\bf{1}} \equiv  \int\limits_{-\infty}^{\infty}\diff y\, V_1\bigl(\hat T_{\bf{4}\mbox{-}\bf{1}}\bigr) = &
 \frac{416 v^3 \sqrt{\lambda} \alpha}{375}\biggl(-11+\frac{6R}{\tanh(R)}-\frac{15}{\sinh^2(R)}\Bigl(1-\frac{R}{\tanh(R)}\Bigr)\biggr)\nonumber \\ & + \frac{16 v^3 \sqrt{\lambda} \mu^2}{5}\Bigl(1-\frac{R}{\tanh(R)}\Bigr)\,.
\end{align}
We see that $V_{\rm eff}^{\bf{3}\mbox{-}\bf{2}}$ is always bellow $V_{\rm eff}^{\bf{4}\mbox{-}\bf{1}}$ (except the point $R=0$). 
We display these effective potentials on Fig.~\ref{fig:effpots}, showing that the minimum is reached at some non-zero separations of 
 coincident triplet and coincident doublet of walls.

 A furhter analysis can be made by considering more complicated configurations, such as $\bf{2}\mbox{-}\bf{1}\mbox{-}\bf{2}$ double-split background. Omitting the details, the effective potential shows that minimum is achieved by collapsing one of the gaps, thus reaching \threetwo configuration.
 In the following section, and throughout the paper, we consider the ${\bf 3}$-${\bf 2}$ configuration as our domain wall background solution.

%
%
\section{Higgs sector}\label{sec:Higgs}
In this section, we propose two models for the Higgs sector, which solve the double-triplet splitting problem.
One model does not need fine-tuning (Model 1), while the other model (Model 2) does.
\subsection{Model 1}
\label{sec:HSWOF}
We introduce a non-minimal kinetic term for the Higgs field ${\cal H}$:
\be
{\cal L}_{\rm H} = \tr \left[\beta_H(\hat T)^2\, {\cal D}_M {\cal H} {\cal D}^M{\cal H}^\dagger \right]\,,
\label{eq:beta_Higgs}
\ee
where ${\cal D}_M {\cal H}=\partial_M {\cal H}-i{\cal A}_{M}{\cal H}$, and we assume $\beta_H$ is transposed as $\beta_H \to U \beta_H U^\dag$ under $SU(5)$.
This is a generalization of Ohta and Sakai idea \cite{Ohta:2010fu} in Eq.~(\ref{eq:OS_mechanism}) to the Higgs sector.
As we will shortly show, no other terms are needed to solve the doublet-triplet splitting problem. 


To be concrete, let us take a specific example
\be
\beta_H(\hat T) = \exp \left(-c\, \hat T^{-1}\right),
\label{eq:beta_H}
\ee
with $c$ being a constant. 
We use this example to demonstrate the idea, but we will comment about the general conditions on $\beta_H$ later.

We consider the ${\bf 3}$-${\bf 2}$ background wall configuration given by
%
\be
T&\equiv&
\begin{pmatrix}
 \tau_3 & 0\\
 0 & \tau_2
\end{pmatrix}
=v \begin{pmatrix}{}
\tanh\Omega\left(y+{d \over 2}\right){\bf 1}_3 & 0 \\
0 & \tanh\Omega\left(y-{d \over 2}\right){\bf 1}_2
\end{pmatrix}\,.
\ee
From this, we have
\be
\hat T &=& \delta \tau\, {\rm diag}\left(\frac{2}{5},\ \frac{2}{5},\ \frac{2}{5},\ -\frac{3}{5},\ -\frac{3}{5}\right),
\label{eq:adjoint_32}\\
\delta \tau &=& v \tanh\Omega\left(y+{d \over 2}\right) - v \tanh\Omega\left(y-{d \over 2}\right) .
\ee
Then, the explicit form of $\beta_H$ reads:
\be
\beta_H = 
\left(
\begin{array}{cc}
\beta_{H,3} {\bf 1}_3 & \\
& \beta_{H,2} {\bf 1}_2
\end{array}
\right)
= 
\left(
\begin{array}{cc}
\exp\left(-\frac{5c}{2\delta \tau}\right) {\bf 1}_3 & \\
& \exp\left(\frac{5c}{3\delta \tau}\right) {\bf 1}_2
\end{array}
\right). \label{eq:betaH}
\ee

\begin{figure}[ht]
\begin{center}
\includegraphics[width=15cm]{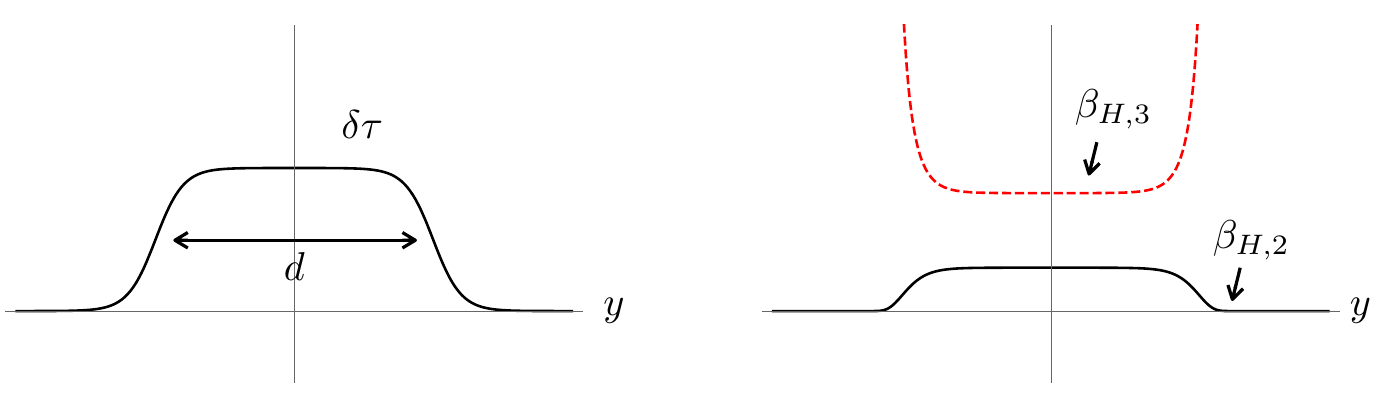}
\caption{\small The left panel shows $\delta\tau$, and the right panel shows $\beta_{H,3}$ and $\beta_{H,2}$.}
\label{fig:bega_H}
\end{center}
\end{figure}

In the following, we assume $c>0$ and $d >0$ (but $c<0$ and $d<0$ would be equally valid). As a consequence, $\beta_{H,3}$ and $\beta_{H,2}$ acquire shapes depicted in Fig.~\ref{fig:bega_H}. 
The sign combination of $c$ and $d$ affects the normalizablity (square integrability) of $\beta_H$. When $cd > 0$ as shown in 
Fig.~\ref{fig:bega_H}, $\beta_{H,2}$ is normalizable but $\beta_{H,3}$ is non normalizable, and vise versa for $cd < 0$.

This naturally leads to decomposing the $5$-plet ${\cal H}$ into the colored triplet ${\cal H}_3$ 
and the electroweak doublet ${\cal H}_2$ as
\begin{eqnarray}
{\cal H}= (\!
\begin{array}{cc}
{\cal H}_3 , & {\cal H}_2 
\end{array}\!
)^T\,,
\label{eq:decomposition_higgs}
\end{eqnarray}
resulting in the decoupled equations of motion
\be
\p_M \left(\beta_{H,I}^2\p^M {\cal H}_I\right) = 0,\qquad (I = 2,3). \label{eq:eomHiggs}
\ee
Since $\beta_{H,I}$ only depends on $y$, this can be expanded as
\be
\beta_{H,I}^2 \p_\mu\p^\mu {\cal H}_I - \p_y\left(\beta_{H,I}^2\p_y {\cal H}_I\right) = 0.
\ee
Next, let us define a canonical Higgs fields $H_I$ defined by
\be
H_I(x^M) = \beta_{H,I}(y) {\cal H}_I(x^M), \label{eq:cn}
\ee
and the equations of motion for $H_I$ read
\be
\p_\mu\p^\mu H_I + \left(-\frac{d^2}{dy^2} + \frac{\beta_{H,I}''}{\beta_{H,I}}\right) H_I = 0.
\ee
Let us expand $H_I$ in to the complete basis of functions
\be
H_I(x^M) = \sum_n f_{I,n}(y)h_{I,n}(x^\mu)\,, \label{eq:he2}
\ee
where $f_{I, n}$ are the eigenfunctions of the operator $-\frac{d^2}{dy^2} + \frac{\beta_{H,I}''}{\beta_{H,I}}$, i.e.,
\be
\left(-\frac{d^2}{dy^2} + \frac{\beta_{H,I}''}{\beta_{H,I}}\right) f_{I,n}(y) = m_{I,n}^2 f_{I,n}(y).
\ee
Clearly, $m_{I,n}$ defines the four-dimensional mass for the $h_{I,n}$.

Let us examine the mass spectra. First of all, we can easily show the mass operator is real and positive semi-definite by
rewriting it in the following form
\be
-\frac{d^2}{dy^2} + \frac{\beta_{H,I}''}{\beta_{H,I}} = Q_{H,I}^\dagger Q_{H,I}, 
\ee
where
\be
Q_{H,I} \equiv -\frac{d}{dy} + (\log \beta_{H,I})',\quad
Q_{H,I}^\dagger \equiv  \frac{d}{dy} + (\log \beta_{H,I})'.
\ee
Thus, the four-dimensional masses satisfy $m_{H,n}^2 \ge 0$. Secondly, irrespective of any details of $\beta_{H,I}$,
the zero mode always exists
\be
Q_{H,I}f_{I,n=0} = 0 \quad \Rightarrow \quad
f_{I,n=0} = N_{H,I}\beta_{H,I}\,. \label{eq:zh}
\ee
where $N_{H,I}$ is a normalization constant which is determined by a condition
\be
 \int dy~f_{I,0}(y)f_{I,0}(y)=1\,.
\ee
This gives
\be
 N_{H,I}={1 \over \sqrt{\int dy \,\beta_{H,I}^2}}\,. \label{eq:norma-fac}
\ee
However, only the electroweak doublet $h_{2,0}$ turns out to be physical, since $\beta_{H,2}$ is only normalizable.
The colored triplet $h_{3,0}$ is unphysical due to the fact that $\beta_{H_3}$ is non-normalizable!
The presence of the massive bound states depends on the details of the model parameters, but they are always
superheavy because all the dimensionful parameters are of the order of the GUT scale.
Thus, the doublet-triplet splitting puzzle is solved without any fine-tuning in this model.

Let us make comments on our accomplishment in this subsection.
The presence of the massless mode is robust in the sense that a small deformation of the model cannot eliminate it.
We can understand it as an analogy to a topological edge mode appearing on an interface between topologically distinct regions. For this, we focus on the mass operator,
which was found to appear in the very specific form $Q_{H,I}^\dagger Q_{H,I}$. This reminds us of the domain wall fermions
via the so-called Jackiw-Rebbi mechanism \cite{Jackiw:1975fn}. Briefly speaking, the Jackiw-Rebbi mechanism states that the massless fermion  appears whenever the position-dependent fermion mass $M_\psi(y)$ changes its sign on a kink center. Typically, the mass appears in the first order 
differential operator $Q_\psi = -\frac{d}{dy} + M_\psi$ and $Q_\psi^\dagger = \frac{d}{dy} + M_\psi$.
Comparing the Jackiw-Rebbi fermion and our Higgs boson, we reach the correspondence 
\be
M_\psi \quad \Leftrightarrow \quad M_{H,I} \equiv (\log\beta_{H,I})' = \frac{\beta_{H,I}'}{\beta_{H,I}}.
\ee
Looking at Fig.~\ref{fig:bega_H}, one sees that $M_{H,I}$ changes the sign at the center.
This, namely the fact that $M_{H,I}(y)$ changes the sign, is a necessary condition for having the massless Higgs doublet.
Any details are not important for the presence of the zero mode. In this sense, the massless Higgs doublet is a sort of the topological edge state. In summary, the ${\bf 3}$-${\bf 2}$ splitting of the 5 domain walls solves the doublet-triplet splitting puzzle of $SU(5)$ GUT.

From the above analysis, if we replace the parameter combination $cd > 0$ by $cd < 0$, we immediately see that
the colored Higgs triplet provides the massless physical mode instead of the doublet. Unfortunately, we cannot control
whether it is the triplet or the doublet that gives the massless degrees of freedom at this stage.
It is our opinion that it is not a high price to pay, compared to some other $SU(5)$ braneworld models, which need parity assignment to all the bulk and brane fields. We do not need any parity assignment, but we need the condition $cd > 0$.

The example \refer{eq:betaH} illustrates the general idea behind the localization mechanism for the Higgs doublet: 
the non-minimal coupling term $\beta_H\bigl(\hat T\bigr)$ depends on the domain-wall background in such a way that distinguishes the doublet from triplet via exponent-type dependence that -- due to the sign difference -- makes the colored triplet part $\beta_{H,3}$ non-normalizable, while localizing the electroweak doublet $\beta_{H,2}$. Let us stress that any type of profile  $\beta_H\bigl(\hat T\bigr)$ that has the same exponent-type sensitivity would produce the same result, making our scheme, in this sense, robust. However, at this point, it seems difficult to formulate general conditions on $\beta_H$ that would guarantee the same results.

Furthermore, there is an obvious question to what degree is the presented construction natural. In this regard, we can only offer the above-mentioned similarity to Jackiw-Rebbi mechanism for fermionic modes, making the Model 1 a bosonic analog for a topological edge-state.  However, we leave the full motivation as a possible future project.

%
%
\subsection{Model 2}\label{sec:HSWF}
In this subsection\footnote{This section was written with significant contributions from the late Professor Norisuke Sakai.}, we propose a model for the Higgs sector that needs fine-tuning.
The Lagrangian is given by
\begin{equation}
{\mathcal L}_{\rm H} = ({\cal D}_M {\cal H})^\dagger {\cal D}^M {\cal H}
 - V^{(4)} - V^{(2)}\,.
\label{eq:H_lag}
\end{equation}
In order to achieve the triplet-doublet splitting, we need to 
use the adjoint scalar field, $\hat T$, to interact with 
${\cal H}$ not via the kinetic terms but via the potential terms. We assume the simplest choice of the potential as 
\begin{equation}
V^{(4)}= \lambda_{\rm H}({\cal H}^\dagger {\cal H})^2, 
\quad 
V^{(2)}=
{\cal H}^\dagger(\Omega_{\rm H}^2{\bf 1}_5+b\hat T){\cal H}, 
\end{equation}
where the quartic coupling $\lambda_H>0$ assures the stability 
of the potential at large values of ${\cal H}$. 
The parameter $\Omega_{\rm H}$ is 
the five-dimensional mass for ${\cal H}$, and $b$ is the coupling 
of the adjoint scalar $\hat T$ with ${\cal H}$. 
The interaction ${\cal H}^\dagger\hat T{\cal H}$ 
with the coupling $b$ is chosen as an operator with the 
lowest mass dimension. 

The adjoint scalar $\hat T$ for the 3-2 split domain wall is given as in Eq.~(\ref{eq:adjoint_32}).
%
%
It again leads to a natural separation of
 the $5$-plet ${\cal H}$ as in (\ref{eq:decomposition_higgs}).
We obtain the decoupled quadratic terms in the potential as 
\begin{equation}
V^{(2)} = U_2(y) {\cal H}_2^\dagger {\cal H}_2 
+  U_3(y) {\cal H}_3^\dagger {\cal H}_3, 
\end{equation}
\begin{equation}
U_2(y) = \Omega_H^2-\frac{U_0}{\cosh^2\Omega y-\tilde d^2 \sinh^2\Omega y}, 
\label{eq:doublet_pot}
\end{equation}
\begin{equation}
U_3(y) = \Omega_H^2+\frac{2}{3}
\frac{U_0}{\cosh^2\Omega y-\tilde d^2 \sinh^2\Omega y}\,,
\end{equation}
where the wall separation is parametrized by a dimensionless separation $\tilde d$ 
\begin{equation}
\tilde d = \tanh \frac{\Omega d}{2}\,.
\label{eq:dimless_separation}
\end{equation}
The strength of the potential $U_0$ is given as 
\begin{equation}
U_0 = \frac{6}{5}\,bv\tanh \frac{\Omega d}{2}\, >\, 0\,. 
\end{equation}
Note that we have chosen the sign of the parameter $b$ 
to make $U_0>0$. 
\begin{figure}[ht]
\begin{center}
\includegraphics[width=10cm]{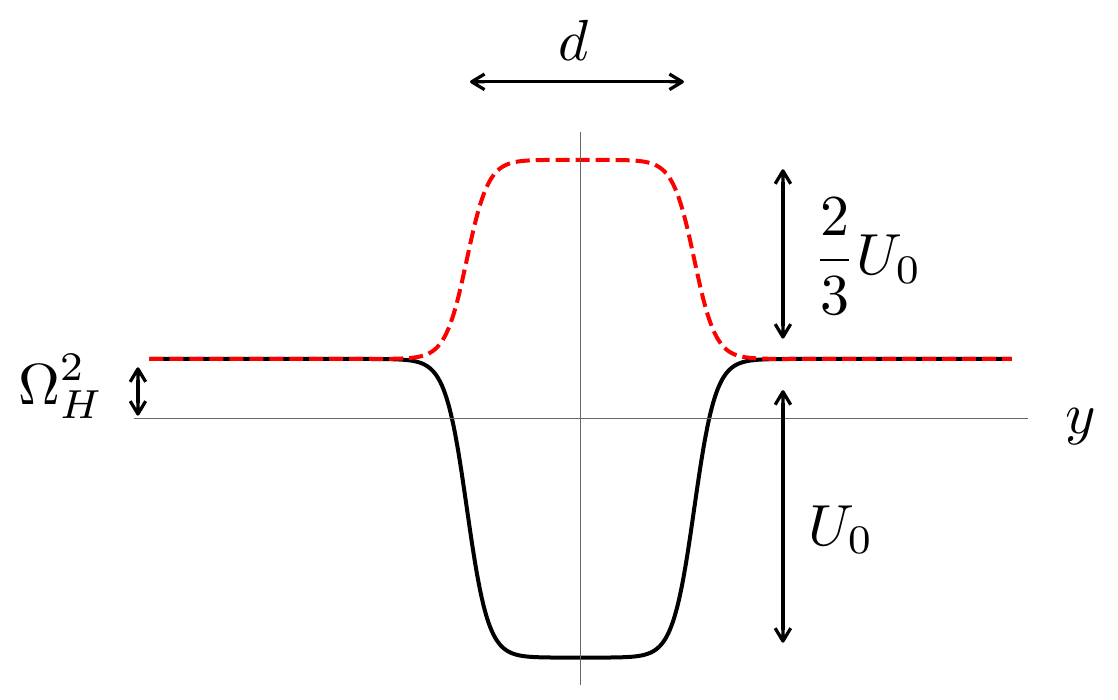}
\caption{\small The quadratic mass terms $U_2$ and $U_3$ for the electroweak 
Higgs doublet (black and solid) and the colored Higgs triplet (red and dashed), respectively.}
\label{fig:V_higgs}
\end{center}
\end{figure}
Because of the repulsive potential for the triplet ${\cal H}_3$, 
there are no localized modes and a large mass gap of order $\Omega_H$ without light modes. 
Therefore we only need to consider doublet ${\cal H}_2$ to 
find possible light or tachyonic modes, see Fig.~\ref{fig:V_higgs}.

The mass spectra of the doublet Higgs scalar ${\cal H}_2$ can be 
obtained by a linearized field equation 
\begin{equation}
{\cal D}_M{\cal D}^M {\cal H}_2+U_2(y){\cal H}_2=0. \label{eq:Higgs-EOM}
\end{equation}
The wave equation in the extra-dimension $y$ is assumed to give 
a complete set of eigenfunctions $u_n(y)$ with the eigenvalue 
$m_n^2$ as 
\begin{equation}\label{eq:eigenU2}
[-\partial_y^2+U_2(y)]u_n(y)=m_n^2 u_n(y).  
\end{equation}
We can decompose the field in terms of this complete set to 
obtain a tower of the Kaluza-Klein fields 
$H_n(x)$, which appear as fields with the mass squared $m_n^2$ 
in the effective four-dimensional field theory 
\begin{equation}
 {\cal H}_2(x,y)=\sum_{n=0}^{\infty} h_{2,n}(x) u_n(y),
\label{eq:KKdecomposition}
\end{equation}
\begin{equation}
\partial_\mu\partial^\mu h_{2,n}(x)+m_n^2 h_{2,n}(x)=0, \label{eq:4dhiggs}
\end{equation}
where $\mu$ stands for the four-dimensional spacetime indices.

It is not easy to obtain the mode functions $u_n(y)$ 
exactly for arbitrary separations. However, since we are 
interested in wall configurations with separations of a 
fraction of the wall width $1/\Omega$, we can solve the eigenproblem \refer{eq:eigenU2}
in the limit of vanishing separation, i.e., $\tilde d\to 0$ as a leading order approximation.

When $\tilde d=0$, the eigenproblem simplifies into
\begin{equation}
\left[-\partial_y^2+\Omega_{\rm H}^2-\frac{U_0}{\cosh^2\Omega y}\right]
u_n(y)=m_n^2 u_n(y).  
\end{equation}
We define an auxiliary variable $s$ 
\begin{equation}
s=\frac{1}{2}\left(\sqrt{1+\frac{4U_0}{\Omega^2}}-1\right), \label{eq:variable}
\end{equation}
to count the discrete eigenstates with the mass $m_n$ 
\begin{equation}
m_n^2 = \Omega_{\rm H}^2-\Omega^2(s-n)^2, \quad n=0, 1, \cdots, \lfloor s\rfloor , 
\end{equation}
where $\lfloor s\rfloor$ is the largest integer less or equal to $s$. 
The corresponding eigenfunctions are given by the  ${}_2F_1$ hypergeometric 
function
\begin{eqnarray}\label{eq:af}
u_n(y) &=& \frac{N_n}{(\cosh \Omega y)^{s-n}} 
{}_2 F_1 \left[-n, 2s-n+1; s-n+1; \frac{1-\tanh \Omega_5 y}{2}\right]
\\
&=& \frac{N_n}{(\cosh \Omega y)^{s-n}} \sum_{k=0}^n 
\frac{n!\Gamma(2s-n+k+1)\Gamma(s-n+1)}{k!(n-k)!\Gamma(2s-n+1)\Gamma(s-n+k+1)}
\left(\frac{\tanh \Omega y}{2}-1\right)^k,  \nonumber 
\end{eqnarray}
with the normalization constant $N_n$ defined by $\int dy |u_n(y)|^2=1$. 
In particular, the lowest mode is given as
\begin{eqnarray}
m_0^2=\Omega_{\rm H}^2-(s\Omega)^2, \quad 
u_0(y) = \frac{N_0}{(\cosh \Omega y)^{s}}, \quad 
|N_0|^2=\frac{\Omega\Gamma(s+\frac{1}{2})}{\sqrt{\pi}\Gamma(s)}. \label{eq:zero-H}
\end{eqnarray}
It is localized between 3 and 2 walls, and it
 becomes massless $m_0=0$ if the parameters satisfy the condition 
\be
U_0 =\Omega_{\rm H}(\Omega_{\rm H}+\Omega). \label{eq:ftc}
\ee 
The mass gap between the lightest and the higher excited modes is of order $\Omega$, and 
above the threshold at $\Omega_{\rm H}^2$, we have a continuum spectra. 

The electroweak symmetry breaking occurs when $m_0^2<0$. 
This is realized when parameters satisfy the the following inequality
\be
 -\sqrt{{\Omega^2 \over 4}+U_0}+{\Omega \over 2}<\Omega_{\rm H}<\sqrt{{\Omega^2 \over 4}+U_0}-{\Omega \over 2}\,.
\ee
We conclude that in this model, the triplet Higgs ${\cal H}_3$ is safely decoupled and the electroweak symmetry breaking is realized by condensation of the remaining double Higgs ${\cal H}_2$.


%
%
\section{Effective Lagrangian}\label{sec:FerYu}
In this section, we give the Lagrangian of the gauge sector, the fermion sector, and the Yukawa coupling in
 \eqref{total}, and derive the four-dimensional effective Lagrangian of \eqref{total}.
\subsection{Effective Lagrangian of gauge sector}\label{sec:gauge_eff}
\subsubsection{Setup}
We consider the following Lagrangian for the gauge sector
\begin{eqnarray}
{\cal L}_{\rm gauge}=-\tr \left [\beta^2 (T_0, \hat{T}) \fcal_{MN} \fcal^{MN} \right]\,, \label{g-k}
\end{eqnarray}
The coefficient $\beta^2 (T_0, \hat{T})$ of the gauge kinetic term, 
which is normally an inverse of a gauge coupling constant 
 but it is a positive semi-definite function of $T_0$ and $\hat{T}$ in the above Lagrangian.
This generalization gives a semi-classical realization of the confinement phase outside the domain wall
 and the localization of the gauge field is achieved.
We require that $\beta^2(T_0, \hat{T})$ covariantly transforms under $SU(5)$ as $\beta_H$ to maintain the gauge invariance.

For the derivation of the four-dimensional effective Lagrangian, we do not need to consider a concrete $\beta(T_0, \hat{T})$. But we only decompose $\beta(T_0, \hat{T})$ in a block-diagonal form, given that the background solution (\ref{eq:dsols}) is block-diagonal as well. That is we have
 \be
 \beta(T_0, \hat{T})|_{\rm bg}=
\begin{pmatrix}
\beta_3(\tau_3)^2{\bf 1}_3 & 0 \\
0 & \beta_2(\tau_2)^2 {\bf 1}_2
\end{pmatrix}\,, \quad \tau_i\equiv v\tanh(\Omega(y-{\cal Y}_i))\quad (i=2, 3)\,.
\ee
As mentioned before,
 the $\beta_3$ and $\beta_2$ realize the localization of the gauge fields by acquiring a profile on the ${\bf 3}$-$\bf{2}$-split background that is approaching zero everywhere, except around the domain walls $\tau_3$ and $\tau_2$, respectively.
Namely we assume square integrability of $\beta_3$ and $\beta_2$ as
\be
{1 \over g_3^2} \equiv \int d y~ \beta_{3}^{2} < \infty\,, \qquad {1 \over g_2^2}\equiv\int d y~ \beta_{2}^{2} < \infty\,. \label{eq:si}
\ee

Let us now decompose the gauge fields as.
\be
 {\cal A}_M =
 \begin{pmatrix}
  {\cal A}_{3M} & {{\cal X}_M \over \sqrt{2}} \\
  {{\cal X}_M^\dagger \over \sqrt{2}} & {\cal A}_{2M}
 \end{pmatrix}
 +{\cal A}_{1M}\sqrt{3 \over 5}
 \begin{pmatrix}
 {1\over 3}{\bf 1}_3 & 0 \\
 0 & -{1 \over 2}{\bf 1}_2
 \end{pmatrix}\,, \label{deco}
\ee
where ${\cal A}_3^M$, ${\cal A}_2^M$, and ${\cal A}_1^M$ are would-be $SU(3)$, $SU(2)$, and $U(1)$ gauge fields, and 
 ${\cal X}_M$ is a $2\times 3$ rectangular complex matrix. 
 
 In addition to the fluctuations of the gauge fields (\ref{deco}), we introduce small fluctuations to the scalar 
 field $T$. 
Since $T$ is a 5 by 5 Hermitian matrix, there are 25 real fluctuations.
We shall decompose them into three parts $(\rho, \Psi, \Phi)$ as
\be
T = e^{i \Phi} (T_0 + \rho) e^{- i \Phi }= T_0 + \rho+i[\Phi, T_0]\cdots\,, \label{eq:fluc-t}
\ee
with
\be
 \rho=
 \begin{pmatrix}
   \rho_3 & 0 \\
   0 & \rho_2
 \end{pmatrix}\,,\quad
\Phi=
{1 \over \sqrt{2}}\begin{pmatrix}
 0 & \varphi \\
 \varphi^\dagger & 0
\end{pmatrix}\,.
\ee
Here $\rho_3\ (\rho_2)$ is a $3\times 3$ ($2\times 2$) Hermitian matrix, and $\varphi$ is $3\times 2$ rectangular, complex matrix.
$\rho_3, \rho_2$ and $\varphi$ have 9, 4 and 12 degrees of freedom.
Summing up, we have 25 degrees of freedom as expected.

The fields $\rho_3$ and $\rho_2$ become massive after the moduli are stabilized as is done in 
 Sec. \ref{sec:moduli_stabilization} and they naturally have GUT scale masses.
Thus, they are integrated out in the low energy Lagrangian.
The third term in the second equality in (\ref{eq:fluc-t}) is nothing but the gauge transformation by the broken generators, so that 
$\varphi$ contains the corresponding Nambu-Goldstone fields.

For later convenience, let us prepare
\be
 \beta_1\equiv \sqrt{{3\beta_2^2 +2\beta_3^2} \over 5}\,,\quad \beta_X\equiv \sqrt{{\beta_2^2+\beta_3^2} \over 2}\,, 
  \quad \beta_\phi\equiv \tau_3-\tau_2\,. \label{eq:beta_phi}
\ee
Here $\beta_1$ also satisfies the square integrability condition since $\beta_3$ and $\beta_2$ satisfy it:
\be
{1 \over g_1^2} \equiv \int d y~ \beta_{1}^{2} < \infty\,. \label{eq:si-1}
\ee

We shall substitute (\ref{deco}) and (\ref{eq:fluc-t}) into (\ref{g-k}) and (\ref{eq:32model}), and pick up only terms of the quadratic 
 order of the fluctuations.
The resultant Lagrangian is then given by
\be
\lcal_{\rm gauge}&+&\tr \left [\dcal_{M} T \dcal^{M} T  \right ]=\sum_{\alpha=1}^3 \lcal_\alpha+\lcal_X+\lcal_\phi \,,
\ee
with
\be
\lcal_\alpha & \equiv & \tr \biggl [ 
\Acal_{\alpha\mu} \left \{\beta_{\alpha}^{2} \left ( \eta^{\mu \nu} \Box - \partial^{\mu} \partial^{\nu} \right ) - \eta^{\mu \nu} \partial_{y} \beta_{\alpha}^{2} \partial_{y} \right \} \Acal_{\alpha\nu} \notag \\
&& - 2 ( \partial^{\mu} \Acal_{\alpha\mu} ) \partial_{y} ( \beta_{\alpha}^{2} \Acal_{\alpha y} )
  - \Acal_{\alpha y} \left ( \beta_{\alpha}^{2} \Box \right ) \Acal_{\alpha y} 
\biggr ]\,, \label{eq:g-d}\\
\lcal_X &=& \tr \biggl [
 \xcal_{\mu}^{\dagger} \left [ \beta_X^2 \left ( \eta^{\mu \nu} \Box - \partial^{\mu} \partial^{\nu} \right ) - \eta^{\mu \nu} \partial_{y} \beta_X^2 \partial_{y} + \eta^{\mu \nu} \beta_\phi^{2} \right ] \xcal_{\nu} \notag \\
&& - 2 ( \partial^{\mu} \xcal_{\mu}^{\dagger} ) \left [ \partial_{y} \left ( \beta_X^2 \xcal_{y} \right ) + \beta_\phi^{2} \varphi \right ] 
- \xcal_{y}^{\dagger} \left ( \beta_X^2 \Box + \beta_\phi^{2} \right ) \xcal_{y}\biggr ]\,, \label{eq:x} \\
\lcal_\phi&=& \tr \biggl [- \varphi^{\dagger} \left (\beta_\phi^{2} \Box - \partial_{y} \beta_\phi^{2} \partial_{y} \right ) \varphi 
+ \varphi^{\dagger} \partial_{y} \left ( \beta_\phi^{2} \xcal_{y} \right ) - \beta_\phi^{2} \xcal_{y}^{\dagger} \partial_{y} \varphi 
\biggl ]\,, \label{eq:varphi}
\ee
where $\lcal_\alpha$ and $\lcal_X$ are the Lagrangians for the unbroken and broken gauge fields, respectively.
The indices $\mu,\nu$ denote the four-dimensional space-time, $y$ is a five-dimensional coordinate, and 
 $\alpha=1,2,3$ represents indices of $U(1)$, $SU(2)$ and $SU(3)$ gauge fields. 
It should be noticed that there are mixing terms between $\xcal_y$ the Nambu-Goldstone field $\varphi$ in (\ref{eq:x})
 and also in (\ref{eq:varphi}).
Note that the trace in (\ref{eq:g-d}) for $\Acal_{1\mu}$ should be understood as replacing ${\rm Tr}$ with ${1 \over 2}$.

Together with the above Lagrangians, we need the following gauge fixing terms:
\be
\lcal_{A}^{\rm (gf)} =\sum_{\alpha=1}^3\lcal_{\alpha A}^{\rm (gf)}\,,\quad
\lcal_{\alpha A}^{\rm (gf)} = -\dfrac{\beta_{\alpha}^{2}}{\xi} \tr \left[
\left \{ \partial^{\mu} \Acal_{\alpha\mu} - \dfrac{\xi}{\beta_{\alpha}^{2}} \partial_{y} (\beta_{\alpha}^{2} \Acal_{\alpha y}) \right \}^{2}
\right]\,,
\ee
As mentioned above $\rm Tr$ is replaced with $1 \over 2$ for $\alpha=1$. We also need another gauge fixing term:
\be
\lcal_{X\phi}^{\rm (gf)}= - \dfrac{\beta_X^{2}}{\xi} \tr \left [ \left\{\partial^{\mu} \xcal_{M} - \dfrac{\xi}{\beta_X^{2}} \left ( \partial_{y} \left ( \beta_X^{2} \xcal_{y} \right ) + \beta_\phi^{2} \varphi \right )\right\}\times ({\rm h.c.}) \right ]\,, \label{eq:gf-xphi}
\ee
where $\xi$ is a gauge fixing parameter similar to a standard $R_\xi $ gauge fixing procedure.

\subsubsection{Mode analysis for the unbroken gauge fields}

The above quadratic Lagrangian has a complicated form because of the existence of the factor $\beta_{\alpha,X}$.
In the following, we will bring the Lagrangian to a standard form. First we shall focus on the unbroken part,
 $\lcal_\alpha$ and the corresponding gauge fixing term $\lcal_{\alpha A}^{\rm (gf)}$.
In the beginning, let us define
\be
 A_{\alpha M}\equiv \beta_\alpha\Acal_M^\alpha\,,\quad (\alpha=1,2,3)\,,  \label{eq:can}
\ee
where the index $\alpha$ is not summed.
We also introduce differential operators
\be
D_{\alpha}\equiv  -\beta_\alpha\partial_y{1 \over \beta_\alpha}=-\partial_y+(\beta_\alpha^{-1}\partial_y\beta_\alpha)\,,\quad (\alpha=1,2,3)\,, \label{eq:dy} 
\ee
where the index $\alpha$ is again not summed.
An adjoint of $D_\alpha$ is given by
\be
 D^\dagger_\alpha=\beta_\alpha^{-1}\partial_y\beta_\alpha=\partial_y+(\beta_\alpha^{-1}\partial_y\beta_\alpha)\,,\quad(\alpha=1,2,3)\,, \label{eq:dya}
\ee
where again we do not sum over $\alpha$. 

Now we are ready to rewrite the effective Lagrangian (\ref{eq:g-d}) by using the above differential operators:
\be
\lcal_\alpha=\tr\left[
A_{\alpha\mu}(\eta^{\mu\nu}\Box-\partial^\mu\partial^\nu+\eta^{\mu\nu}\triangle_\alpha)A_{\alpha\nu}
-2(\partial^\mu A_\mu)(D_\alpha^\dagger A_{\alpha y})
-A_{\alpha y}\Box A_{\alpha y}
\right]\,,
\ee
with $\triangle_\alpha\equiv D_\alpha^\dagger D_\alpha$ and $\square = \p_\mu\p^\mu$.
Next we rewrite the gauge fixing Lagrangian as
\be
\lcal_{\alpha A}^{\rm (gf)}=-{1 \over \xi}\tr\left[\left(\partial^\mu A_{\alpha\mu}-\xi D^\dagger_\alpha A_{\alpha y} \right) \right]^2\,.
\ee
Summing up these, we obtain relatively simple formula for the unbroken gauge fields
\be
\lcal_{\rm ub}\equiv\sum_{\alpha=1}^3(\lcal_\alpha+\lcal_{\alpha A}^{\rm gf})&=&\sum_{\alpha=1}^3\tr\Bigg [
A_{\alpha\mu}\left(\eta^{\mu\nu}\Box-\left(1-{1 \over \xi}\right)\partial^\mu\partial^\nu+\eta^{\mu\nu}\triangle_\alpha\right)A_{\alpha\nu} \nonumber \\
&&-A_{\alpha y}(\Box+\xi D_\alpha D_\alpha^\dagger) A_{\alpha y}\Bigg ]\,, \label{eq:sum-A}
\ee
where we have used
\be
 (D_\alpha^\dagger A_{\alpha y})D_\alpha^\dagger A_{\alpha y}=A_{\alpha y}D_\alpha D_\alpha^\dagger A_{\alpha y}\,,
\ee
where we do not sum over $\alpha$. 

It is possible to decompose the extra-dimensional gauge field $A_{\alpha y}$ into two scalar fields $B_\alpha$ 
 and $C_\alpha$ as
\be
A_{\alpha y}=D_\alpha B_\alpha + C_\alpha\,, \label{eq:deco-A}
\ee
where $C_\alpha$ satisfies the divergence-free condition
\be
D^\dagger_\alpha C_\alpha=0\,. \label{eq:div}
\ee
We prove this  decomposition in \ref{app:B}.

Multiplying $D_\alpha^\dag$ on (\ref{eq:deco-A}) leads to the following equation
\be
 \triangle_\alpha  B_\alpha = D^\dagger_\alpha A_{\alpha y}\,. \label{eq:BA}
\ee
We can determine $B_\alpha$ and $C_\alpha$ by solving the equations (\ref{eq:div}) and (\ref{eq:BA}).
It is easy to solve Eq. (\ref{eq:div}) as
\be
 C_\alpha \propto\beta_\alpha^{-1}\,.
\ee
From the square integrability conditions (\ref{eq:si}) and (\ref{eq:si-1}), we see that $C_\alpha$ is 
 non-normalizable.
The scalar $B_\alpha$ can be derived for a given $A_{\alpha y}$ in (\ref{eq:BA}). 
However, there is an ambiguity. 
We can determine $B_\alpha$ only up to solutions of the following equation
\be
 \triangle_\alpha  B^{(0)}_{\alpha}=0\,. \label{eq:lap}
\ee
A solution to this equation is the kernel of $D_\alpha$, which is given by
\be
 B^{(0)}_{\alpha}\propto \beta_\alpha\,. \label{eq:solB}
\ee
Note that this is normalizable from the square integrability conditions (\ref{eq:si}) and (\ref{eq:si-1}).

By using the decomposition (\ref{eq:deco-A}) we can rewrite (\ref{eq:sum-A}) as
\be
\lcal_{\rm ub}&=&\sum_{\alpha=1}^3\tr\Bigg [
A_{\alpha\mu}\left(\eta^{\mu\nu}\Box-\left(1-{1 \over \xi}\right)\partial^\mu\partial^\nu+\eta^{\mu\nu}\triangle_\alpha\right)A_{\alpha\nu} \nonumber \\
&&
-C_\alpha\Box C_\alpha-B_{\alpha}\triangle_\alpha(\Box+\xi \triangle_\alpha) B_{\alpha }\Bigg ]\,. \label{eq:decom-A2}
\ee
We redefine the scalar $B_\alpha$ by
\be
 b_\alpha \equiv \sqrt{\triangle_\alpha}B_\alpha\,. \label{eq:Delta}
\ee
Substituting this into (\ref{eq:decom-A2}), we have 
\be
\lcal_{\rm ub}=\sum_{\alpha=1}^3\tr\Bigg [
A_{\alpha\mu}\left(\eta^{\mu\nu}\Box-\left(1-{1 \over \xi}\right)\partial^\mu\partial^\nu+\eta^{\mu\nu}\triangle_\alpha\right)A_{\alpha\nu}-b_{\alpha}(\Box+\xi \triangle_\alpha) b_{\alpha }\Bigg ]\,, \label{eq:decom-A3}
\ee
where $C_\alpha$ are dropped since they are non-normalizable.

We should emphasize that $b_\alpha$ does not include the zero mode. 
It is found that the zero mode for $B_\alpha$ is obtained from $\triangle_\alpha B_\alpha=0$ (see (\ref{eq:decom-A2})). 
This is nothing but (\ref{eq:lap}) and its solution is given by (\ref{eq:solB}). 
However, the zero mode is projected out by the definition (\ref{eq:Delta}) and therefore it does not exist in $b_\alpha$.
Now we decompose $b_\alpha$ by the eigenfunctions of $\triangle_\alpha$
%
\be
 b_\alpha(x^M)=\sum_{n>0}b_\alpha^{(n)}(x^\mu)B_\alpha^{(n)}(y)\,,  \label{eq:B}
 \ee 
where $b^{(n)}_\alpha=f_{\alpha n}m_{\alpha n}$ with $f_{\alpha n}$ being an expansion coefficient 
 (a field in four dimensions) and $m_{\alpha n}$ being the four-dimensional masses.
The mode functions $B_\alpha^{(n)}(y)$ satisfy
\be
 \triangle_\alpha B_\alpha^{(n)}=m_{\alpha n}^2B_\alpha^{(n)}\,, \label{eq:mode-B}
\ee
Note that $n=0$ is not included in the expansions because of the explanation above.
We set the normalization of the mode functions as usual:
\be
 \int dy ~B_{\alpha}^{(m)}B_{\alpha}^{(n)}=\delta^{mn}\,. \label{eq:n}
\ee
%
%
%

In contrast to the extra-dimensional gauge fields, there exists the zero mode of $\triangle_\alpha$ for
the four-dimensional gauge fields. 
Therefore, when we decompose $A_{\alpha \mu}$ in terms of the eigenfunctions of $\triangle_\alpha $, we have to include $n=0$ as
\be
 A_{\alpha \mu}(x^M)=\sum_{n \ge 0}A_{\alpha \mu}^{(n)}(x^\mu)B_{\alpha}^{(n)}(y)\,. \label{eq:decomp-A}
\ee
The mode function for the zero mode is given by (\ref{eq:solB}). 
They are localized around the domain walls, that is the simple and robust resolution of the long standing problem in the brane-world models. 
Note also that the domain walls work as trigger of spontaneous symmetry breaking from GUT to SM symmetries.
The normalization is fixed by (\ref{eq:si}) and (\ref{eq:si-1}) and we have
\be
 B_{\alpha}^{(0)}=g_\alpha \beta_\alpha\,, \label{eq:B-N}
\ee

Substituting (\ref{eq:B}) and (\ref{eq:decomp-A}) into (\ref{eq:decom-A3}) and integrating it over $y$, 
 we obtain the effective Lagrangian for the Kaluza-Klein (KK) towers as
\be
\lcal_{\rm ub}&=&\sum_{n\ge 0}\tr\Bigg [A_{\alpha\mu}^{(n)}\left(\eta^{\mu\nu}\Box-\left(1-{1 \over \xi}\right)
\partial^\mu\partial^\nu+\eta^{\mu\nu}(m_{\alpha n})^2 \right)A^{(n)}_{\alpha\nu} \nonumber \\
&&-\sum_{n>0}b_\alpha^{(n)}(\Box+\xi (m_{\alpha n})^2)b_\alpha^{(n)} \Bigg ]\,.
\ee
Note that the scalar $b_\alpha$ is coming from the extra-dimensional component of $A_{\alpha y}$ and it appears
 with the mass squared  $\xi (m_{\alpha n})^2$. This is nothing but the standard Lagrangian in the so-called $R_\xi$ gauge.
This means that the scalar $b_\alpha$ plays a role as a Nambu-Goldstone boson to give a mass to the KK gauge fields
 through the Higgs mechanism.

Since the KK gauge fields $A_{\alpha \mu}^{(n)}$ for $n>0$ become super-massive, they are integrated out and the effective 
 Lagrangian for the zero modes is given by 
\be
\lcal_{\rm ub}^{\rm eff}=\sum_{\alpha=1}^3\tr\left [
A_{\alpha\mu}^{(0)}\left(\eta^{\mu\nu}\Box-\left(1-{1 \over \xi}\right)\partial^\mu\partial^\nu\right)A^{(0)}_{\alpha\nu}\right]\,.
\ee
This is the quadratic Lagrangian of massless gauge fields. 
In order to recover the self-interaction of the non-Abelian gauge fields, we go back to (\ref{eq:can}).
\be
 \Acal_{\alpha \mu}=\beta_\alpha^{-1}A_{\alpha \mu}=g_\alpha A_{\alpha \mu}^{(0)}+\cdots\,, \label{eq:or}
\ee
where the dots denote the massive modes.
Keeping only the zero mode in (\ref{eq:or}), the field strength corresponding to $\Acal_{\alpha \mu}$ becomes
\be
 {\cal F}_{\alpha\mu\nu}&=&g_\alpha F_{\alpha\mu\nu}^{(0)}\,,
\ee
where 
\be
 F_{\alpha \mu\nu}^{(0)}&=&\partial_\mu A_{\alpha\nu}^{(0)}-\partial_\nu A_{\alpha\mu}^{(0)}+ig_\alpha[A_{\alpha\mu}^{(0)},A_{\alpha\nu}^{(0)}]\,. \label{eq:fs}
\ee
This should be compared with the original field strength 
 $\fcal_{MN} = \partial_{M} \acal_{N} - \partial_{N} \acal_{M} + i [ \acal_{M} , \acal_{N}]$ where the gauge coupling
 is absorbed in the gauge field $\Acal_{M}$.
On the other hand, the four-dimensional gauge coupling appears in (\ref{eq:fs}) as a consequence of integrating
 over the extra-dimensional coordinate $y$.
As a result, the effective Lagrangian for the zero modes including the self-interaction is given by
\be
 \lcal_{\rm ub}^{\rm eff}&=&\int dy \sum_{\alpha=1}^3 \tr \left[-{1 \over 2}\beta_\alpha^2 \fcal_{\alpha \mu\nu}\fcal_{\alpha}^{\mu\nu}\right] \nonumber \\
 &=&-{1 \over 2}{\rm Tr}(G_{\mu\nu}G^{\mu\nu})
  -{1 \over 2}{\rm Tr}(W_{\mu\nu}W^{\mu\nu})
  -{1 \over 4}B_{\mu\nu}B^{\mu\nu}\,,
\ee
where we have defined
 $A_{\alpha\mu}^{(0)}(x)=(A_{1\mu}^{(0)},A_{2\mu}^{(0)},A_{3\mu}^{(0)})\equiv (B_\mu, W_\mu, G_\mu)$ being
 the four-dimensional massless gauge fields for $U(1), SU(2), SU(3)$ gauge groups and corresponding
 field strength
 \be
 G^{\mu\nu}&=&\partial^\mu G^{\nu}-\partial^\nu G^{\mu}+ig_3[G^{\mu},G^{\nu}]\,, \\
 W^{\mu\nu}&=&\partial^\mu W^{\nu}-\partial^\nu W^{\mu}+ig_2[W^{\mu},W^{\nu}]\,, \\
 B^{\mu\nu}&=&\partial^\mu B^{\nu}-\partial^\nu B^{\mu}\,.
\ee
%
%
%
\subsubsection{Mode analysis for the broken gauge fields}
In this subsection, we consider the effective Lagrangian of $\lcal_X$ and $\lcal_\phi$.
First we prepare canonical fields as
\be
 X_M \equiv \beta_X\xcal_M\,,\quad X_5\equiv \beta_\phi \varphi\,.
\ee
It is convenient to define a two-component vector containing the extra-dimensional components of 
 the unbroken gauge field and the Nambu-Goldstone field as
\be
\boldsymbol{X} \equiv 
\begin{pmatrix}
X_4 \\ 
X_5
\end{pmatrix}\,, \label{eq:vecX}
\ee
where we have denoted $X_4\equiv X_y$.
As in the unbroken gauge part, we introduce differential operators
\be
D_X&\equiv & -\beta_X\partial_y{1 \over \beta_X}=-\partial_y+(\beta_X^{-1}\partial_y\beta_X)\,, \label{eq:DX}\\
D_\phi&\equiv &-\beta_\phi\partial_y{1 \over \beta_\phi}=-\partial_y+(\beta_\phi^{-1}\partial_y\beta_\phi)\,. \label{eq:phi}
\ee
Note that $D_X$ can be expressed via $D_\phi$ as
\be
 D_X={\cal M}^{-1}D_\phi{\cal M}\,,\qquad {\cal M}\equiv {\beta_\phi \over \beta_X}\,.
 \label{eq:rel-x}
\ee
Then, we define a two-component operator as
\be
 \boldsymbol{D} \equiv 
\begin{pmatrix}
D_X \\ {\cal M}
\end{pmatrix}\,, \label{eq:Dvec}
\ee
and
\be
 \boldsymbol{D}_R=({\cal M}, -D_\phi)\,. \label{eq:D-rot}
\ee
With the use of (\ref{eq:vecX}), (\ref{eq:DX}), (\ref{eq:phi}) and (\ref{eq:Dvec}), we can rewrite the sum of (\ref{eq:x}) and 
 (\ref{eq:varphi}) in a compact form
\be
\lcal_X+\lcal_{\phi} = 
\tr\Big[X_\mu^\dagger(\eta^{\mu\nu}\Box-\partial^\mu\partial^\nu+\eta^{\mu\nu}\Delta)X_\nu 
-2(\partial^\mu X^\dagger_\mu) \boldsymbol{D}^\dagger \boldsymbol{X}
  -\boldsymbol{X}^\dagger(\Box+\tilde{\Delta})\boldsymbol{X}\Big ]\,, \label{eq:x-phi}
\ee 
where 
\be
&\Delta=\boldsymbol{D}^\dagger\boldsymbol{D}\,, &\\
& \tilde{\Delta}=\boldsymbol{D}_R^\dagger \boldsymbol{D}_R=\begin{pmatrix}
{\cal M}^2 & -{\cal M}D_\phi \\
-D_\phi^\dagger {\cal M} & D_\phi^\dagger D_\phi
\end{pmatrix}\,.&
\ee
Similarly, the gauge fixing Lagrangian (\ref{eq:gf-xphi}) is expressed as
\be
\lcal_{X\phi}^{(\rm gf)}=-{1 \over \xi}\tr\left[(\partial^\nu X_\nu-\xi \boldsymbol{D}^\dagger \boldsymbol{X})^\dagger
(\partial^\mu X_\mu-\xi \boldsymbol{D}^\dagger \boldsymbol{X})\right]\,. \label{eq:gf-x-phi}
\ee
Summing up (\ref{eq:x-phi}) and (\ref{eq:gf-x-phi}), we have the following quadratic Lagrangian
\be
\lcal_{\rm b} &\equiv& \lcal_X+\lcal_{\phi}+\lcal_{X\phi}^{(\rm gf)}  \nonumber\\
&=&
\tr\Bigg [ X_\mu^\dagger(\eta^{\mu\nu}\Box-\left(1-{1 \over \xi}\right)\partial^\mu\partial^\nu+\eta^{\mu\nu}\Delta)X_\nu 
-\boldsymbol{X}^\dagger(\Box+\tilde{\Delta}+\xi \boldsymbol{D}\boldsymbol{D}^\dagger)\boldsymbol{X}\Bigg ]\,, \label{eq:lag-x-phi-1}
\ee
where we have used 
\be
 (\boldsymbol{D}\boldsymbol{X})^\dagger\boldsymbol{D}^\dagger\boldsymbol{X}=\boldsymbol{X}^\dagger \boldsymbol{D}\boldsymbol{D}^\dagger \boldsymbol{X}\,.
\ee

As in the case of the extra-dimensional component of the unbroken gauge fields in Eq.~(\ref{eq:deco-A}), we can show that $\boldsymbol{X}$  
 is decomposed by a real scalar $Y$ and a two-component complex vector $\boldsymbol{Z}$ as
\be
 \boldsymbol{X}=\boldsymbol{D}Y+\boldsymbol{Z}\,, \label{eq:decom-X}
\ee
where $Y$ and $\boldsymbol{Z}$ satisfy
\be
 &\boldsymbol{D}_R\boldsymbol{D}Y=0\,,& \\
 &\boldsymbol{D}^\dagger\boldsymbol{Z}=0\,. \label{eq:Z1}&
\ee
The proof of the decomposition (\ref{eq:decom-X}) is given in \ref{app:B}.
Acting $\boldsymbol{D}^\dagger$ on (\ref{eq:decom-X}), we find
\be
 \boldsymbol{D}^\dagger \boldsymbol{D}Y=\boldsymbol{D}^\dagger \boldsymbol{X}\,, \label{eq:Y1}
\ee
by which $Y$ is determined.
Note that we shall exclude the kernel of $\boldsymbol{D}$ from $Y$ 
 because it cannot participate in the physical field $\boldsymbol{X}$.
Substituting (\ref{eq:decom-X}) into (\ref{eq:lag-x-phi-1}), we have
\be
\lcal_{\rm b}=
\tr\Bigg [ X_\mu^\dagger \left\{ \eta^{\mu\nu}\Box-\left(1-{1 \over \xi}\right)\partial^\mu\partial^\nu+\eta^{\mu\nu}\Delta \right\} X_\nu 
-Y\Delta(\Box+\xi \Delta)Y-\boldsymbol{Z}^\dagger(\Box+\tilde{\Delta})\boldsymbol{Z}\Bigg ]\,. \label{eq:lag-x-phi-2}
\ee
We make this Lagrangian canonical by redefining $Y$ as
\be
 \mathcal{Y}=\sqrt{\Delta}Y\,.
\ee
We excluded the kernel of $\boldsymbol{D}$ from $Y$ and so it is true for $ \mathcal{Y}$. 
Substituting this into (\ref{eq:lag-x-phi-2}), we have the final form of the quadratic Lagrangian of the broken sector
\be
\lcal_{\rm b} =
\tr\Bigg [ X_\mu^\dagger \left\{\eta^{\mu\nu}\Box-\left(1-{1 \over \xi}\right)\partial^\mu\partial^\nu+\eta^{\mu\nu}\Delta \right\} X_\nu 
- \mathcal{Y}(\Box+\xi \Delta) \mathcal{Y}-\boldsymbol{Z}^\dagger(\Box+\tilde{\Delta})\boldsymbol{Z}\Bigg ]\,. \label{eq:lag-x-phi-}
\ee

We shall discuss mass spectra of $X_\mu,  \mathcal{Y}$ and $\boldsymbol{Z}$.
The mass spectrum of $X_\mu$ is controlled by the operator $\Delta= \boldsymbol{D}^\dagger\boldsymbol{D} = D_X^\dagger D_X + {\cal M}^2$.
This means that there is no state satisfying $\Delta X_\mu=0$ and therefore there are no
 massless modes in $X_\mu$.
This is, of course, expected since $X_\mu$ corresponds to the gauge fields associated with the broken 
 generators.
Indeed, $X_\mu$ becomes massive by absorbing $ \mathcal{Y}$.
We are lead to this conclusion by the fact that $X_\mu$ and $ \mathcal{Y}$ share the same mass operator $\Delta$ except 
 for the extra constant factor $\xi$ for $ \mathcal{Y}$.
Again we come across a generalized form of the $R_\xi$ gauge.
This means that $ \mathcal{Y}$ plays a role of infinite tower of Nambu-Goldstone bosons and they are eaten by infinite tower
 of the KK modes of $X_\mu$ including the bottom of the tower.
 
Recalling the definition (\ref{eq:Y1}), the source for $ \mathcal{Y}$ is 
 $\boldsymbol{D}^\dagger \boldsymbol{X}=D_X^\dagger X_4+{\cal M}\beta_\phi\varphi$.
Therefore $X_\mu$ eats not pure Nambu-Goldstone field $\varphi$ but the mixture of $D_X^\dagger X_4$ and $\varphi$.
It is in general difficult to specify the lowest mass of $\Delta$ except for the special case, where $\cal M$ is a constant.
In this case, since $\Delta=D_X^\dagger D_X +{\cal M}^2$, 
$\Delta$ differs from $D_X^\dagger D_X$ by just constant ${\cal M}^2$.
The zero mode of $D_X^\dagger D_X$ is given by $\beta_X$ and therefore the lowest mass of $\Delta$ is the constant ${\cal M}$
 itself.
So if we set ${\cal M}$ to GUT scale, the broken gauge field $X_\mu$ is superheavy.

Let us turn to the mass spectra of $\boldsymbol{Z}$. Note that $\tilde{\Delta}$ is positive semi-definite because
 $\boldsymbol{Z}^\dagger \tilde{\Delta}\boldsymbol{Z}=\boldsymbol{Z}^\dagger \boldsymbol{D}_R^\dagger \boldsymbol{D}_R \boldsymbol{Z}=(\boldsymbol{D}_R\boldsymbol{Z})^\dagger \boldsymbol{D}_R\boldsymbol{Z}$.
The zero modes of $\tilde{\Delta}$ are kernel of $\boldsymbol{D}_R$, but one can show that $\boldsymbol{Z}$ 
 does not include the zero mode.
It is proved in \ref{app:B}. 

In summary, $X_\mu,  \mathcal{Y}$ and $\boldsymbol{Z}$ do not have zero modes and their KK towers start from modes 
 with masses. 
These modes are integrated out at low energies so that we can conclude that modes of $X_\mu,  \mathcal{Y}$ and
 $\boldsymbol{Z}$ do not appear in the low-energy effective Lagrangian.

%
%
\subsection{Effective Lagrangian of fermion sector 1}\label{sec:fer_eff}
In this subsection, we give the Lagrangian of the fermion sector and derive its effective Lagrangian.
The Lagrangian of the fermion part ${\cal L}_{\rm f}$ in (\ref{total}) is written as
\begin{eqnarray}
 {\cal L}_{\rm f}&=&i\bar{\Psi}_{\bar{5}A}\Gamma^MD_M\Psi_{\bar{5}A}+i{\rm Tr}[\bar{\Psi}_{10A}\Gamma^MD_M\Psi_{10A}] \nonumber \\
  && +h_{5A}\bar{\Psi}_{\bar{5}A}\hat{T}^t\Psi_{\bar{5}A}+\tilde{h}_{5A}T^0\bar{\Psi}_{\bar{5}A}  \Psi_{\bar{5}A} \nonumber \\
  && +h_{10A}{\rm Tr}[\bar{\Psi}_{10A}\hat{T}\Psi_{10A}]+\tilde{h}_{10A}T^0{\rm Tr}[\bar{\Psi}_{10A}\Psi_{10A}]\,, \label{eq:fermion}
\end{eqnarray}
where $h_{5A}, \tilde{h}_{5A}, h_{10A}$ and $\tilde{h}_{10A}$ are Yukawa-type coupling constants with $\hat T$ and $T^0$, $A=1,2,3$ are the generation index, and
 $\Gamma^M$ is the five-dimensional gamma matrices defined as 
\begin{eqnarray}
 \Gamma^\mu=\gamma^\mu,\quad \Gamma^5=i\gamma^5,\quad \gamma^5=i\gamma^0\gamma^1\gamma^2\gamma^3\,,
\end{eqnarray}
where $\gamma^\mu$ are the four-dimensional gamma matrices.
The Dirac conjugate is defined by $\bar{\Psi}=\Psi^\dagger\Gamma^0$. 
The covariant derivatives are
\begin{eqnarray}
 D_M\Psi_{\bar{5}A}&=&\partial_M\Psi_{\bar{5}A}-i{\cal A}_M^*\Psi_{\bar{5}A}\,, \\
 D_M\Psi_{10A}&=&\partial_M\Psi_{10A}+i{\cal A}_M\Psi_{10A}+i\Psi_{10A}{\cal A}_M^t\,.
\end{eqnarray}

To derive the effective Lagrangian for ${\cal L}_{\rm f}$ we first decompose the $SU(5)$ representations as $\bar{5}\rightarrow (\bar{3},1)_{1/3}\oplus (1,2)_{-1/2}$ and
 $10\rightarrow (\bar{3},1)_{-2/3}\oplus (3,2)_{1/6}\oplus (1,1)_1$ under $SU(3)\times SU(2)\times U(1)$:
\be
\Psi_{\bar{5}}=
\begin{pmatrix}
 \Psi_{\bar{3}} \\
 \Psi_{2}
\end{pmatrix}\,,\quad
\Psi_{10}={1 \over \sqrt{2}}
\begin{pmatrix}
 \Psi_{\bar{3}^\prime} & \Psi_{3,2} \\
 -\Psi_{3,2}^t & \Psi_1
\end{pmatrix}\,. \label{eq:comp-f}
\ee
Hereafter we will suppress the generation index $A$ for ease of notation.
Next step is expanding the fermions into Kaluza-Klein modes by which the Lagrangian ${\cal L}_{\rm f}$ is diagonalized.
For $\Psi_{\bar{3}}$, we have
\be
 \Psi_{\bar{3}}=\sum_n[f_{\bar{3},L}^{(n)}(y)\psi_{\bar{3},L}^{(n)}(x)+f_{\bar{3},R}^{(n)}(y)\psi_{\bar{3},R}^{(n)}(x)]\,, \label{eq:expand-f}
\ee
where $f_{\bar{3},L}^{(n)}(y)$ and $f_{\bar{3},R}^{(n)}(y)$ are wave functions satisfying the mode equation
\be
 Q_{\bar{3}}^\dagger Q_{\bar{3}} f_{\bar{3}}^{(n)}(y)=(m_{\bar{3}}^{(n)})^2 f_{\bar{3}}^{(n)}(y)\,, \label{eq:mode-eq-f}
\ee
where 
\be
 Q_{\bar{3}}=\partial_y+h_{\bar{3}}(y)\,,\quad Q_{\bar{3}}^\dagger=-\partial_y+h_{\bar{3}}(y)\,,\quad h_{\bar{3}}(y)={2 \over 5}h_5\beta_\phi(y)+\tilde{h}_5\tau_+(y)\,.
\ee
Here $\beta_\phi$ is defined in (\ref{eq:beta_phi}), $\tau_+\equiv 3\tau_3+2\tau_2$.
The asymptotic behaviors of  $y$-dependent Yukawa coupling $h_{\bar 3}(y)$ are $h_{\bar 3}(y=+\infty) = 5 v\tilde h_5$ and $h_{\bar 3}(y=-\infty) = -5 v\tilde h_5$. The most important fact is that the fermion mass gap is opened at $y=\pm \infty$ whereas it is closed at somewhere between $y=-\infty$  and $+\infty$.
$m_{\bar{3}}^{(n)}$ are the masses of the four-dimensional chiral fermions $\psi_{\bar{3},L}^{(n)}, \psi_{\bar{3},R}^{(n)}$
which satisfy
%
\be
 \gamma^5 \psi_{\bar{3},L}^{(n)}=-\psi_{\bar{3},L}^{(n)}\,,\quad \gamma^5 \psi_{\bar{3},R}^{(n)}=\psi_{\bar{3},R}^{(n)}\,. \label{eq:chirality}
\ee
As it turns out, for appropriate choice of parameters $h_5$ and $\tilde{h}_5$, the normalizable zero modes appear only for the
 left-handed components as
\be
 f_{\bar{3},L}^{(0)}&=&\tilde{N}_{\bar{3}}\exp\left(-\int^y h_{\bar{3}}(y')~ dy'\right)\,,  \label{eq:zero1}
\ee
where $\tilde{N}_{\bar{3}}$ is the normalization constant, which is determined by a normalization condition
\be
 \int dy\,(f_{\bar{3},L}^{(0)})^2=1\,. \label{eq:nc}
\ee
We identify the zero modes of the chiral fermion $\psi_{\bar{3},L}^{(n)}$ with the down quarks as
\be
 \Psi_{\bar 3} \supset  \psi_{\bar{3},L}^{(0)}\rightarrow d_a=(d_1^c,d_2^c,d_3^c)\,,
\ee
where $a=1,2,3$ is the $SU(3)$ index and where the superscript ${}^{c}$ denotes charge conjugation.
The same analysis is done for $\Psi_2,~\Psi_3^\prime, \Psi_{3,2}$ and $\Psi_1$ similarly.
The zero modes for them are given by
\begin{eqnarray}
f_{2,L}^{(0)}&=&\tilde{N}_2\exp\left(-\int^y h_{2}(y')~ dy'\right)\,, \quad h_2=-{3 \over 5}h_5\beta_\phi+\tilde{h}_5\tau_+\,, \label{eq:f2L}\\
f_{\bar{3}^\prime,L}^{(0)}&=&\tilde{N}_{\bar{3}^\prime}\exp\left(-\int^y h_{\bar{3}^\prime}(y')~dy'\right)\,, \quad h_{\bar{3}^\prime}={2 \over 5}h_{10}\beta_\phi+\tilde{h}_{10} \tau_+\,, \label{eq:f3primeL}\\
f_{3,2,L}^{(0)}&=&\tilde{N}_{3,2}\exp\left(-\int^y h_{3,2}(y')~dy'\right)\,, \quad h_{3,2}=-{1 \over 10}h_{10}\beta_\phi+\tilde{h}_5\tau_+\,, \label{eq:f32L}\\
f_{1,L}^{(0)}&=&\tilde{N}_1\exp\left(-\int^y h_{1}(y')~dy'\right)\,, \quad h_1=-{3 \over 5}h_{10}\beta_\phi+\tilde{h}_5\tau_+\,, \label{eq:f1L}
\end{eqnarray}
respectively.
All these are normalizable due to the fact that the $y$-dependent Yukawa coupling $h$'s have opposite signs at $y = - \infty$ and $+\infty$, and therefore they are physical massless fermions.
Then, the other quarks and leptons are identified as
\be
& \Psi_2 \supset \psi_{2,L}^{(0)}\rightarrow l_a=\left(\begin{array}{c}e^-\\-\nu_e\end{array}\right)\,, \\
& \Psi_{\bar 3'} \supset  \psi_{\bar{3}^\prime,L}^{(0)}\rightarrow U_{\alpha\beta}^c=\epsilon_{\alpha\beta\gamma}u_\gamma^c=
 \begin{pmatrix}
  0 & u^c_3 & -u^c_2 \\
  -u^c_3 & 0 & u^c_1 \\
  u^c_2 & -u^c_1 & 0
 \end{pmatrix}\,, & \\
& \quad \Psi_{3,2} \supset  \psi_{3,2,L}^{(0)}\rightarrow 
q_{\alpha a}=\begin{pmatrix}
u_1 & d_1 \\
u_2 & d_2 \\
u_3 & d_3 
\end{pmatrix}\,,
\quad
\Psi_1 \supset \psi_{1,L}^{(0)}\rightarrow E^+_{ab}=
 \begin{pmatrix}
 0 & e^+ \\
 -e^+ & 0
 \end{pmatrix}\,. &
\ee
Note that the generation index $A$ is suppressed as we mentioned above.
\footnote{The mass dimensions: $[\text{5D fermion}] = 2$, $[\text{5D Yukawa}] = -1/2$, $[\text{fermion zero mode function)}] = 1/2$,
$[\text{4D fermion}] = 3/2$, and $[\text{4D Yukawa}] = 0$.}

Let us now derive the four-dimensional effective action.
We treat $T$ as the background and ignore its fluctuations because they are higher order corrections.
Substituting the domain wall background
\begin{eqnarray}
 \hat{T}=\begin{pmatrix}
 {2 \over 5}\beta_\phi {\bf 1}_3 & 0 \\
 0 & -{3 \over 5}\beta_\phi{\bf 1}_2
 \end{pmatrix}\,,\quad
 T^0=\tau_+\,.
\end{eqnarray}
and (\ref{eq:expand-f}) into the Lagrangian, 
 picking up the left-handed zero modes, and  integrating over $y$, we find
\be
&& i\int dy\, \bar{\Psi}_{\bar{3}}(\Gamma^M D_M \Psi_{\bar{3}} -ih_{\bar{3}}\Psi_{\bar{3}}) \nonumber \\
&& \quad \rightarrow  i\int dy\,f_{\bar{3},L}^{(0)}(y)\bar{\psi}_{\bar{3},L}^{(0)}(x) \gamma^\mu 
 \left(\partial_\mu-ig_3 A_{3\mu}^{(0)*} -i{1 \over 3}\sqrt{3 \over 5}g_1A_{1\mu}^{(0)*} \right)
 f_{\bar{3},L}^{(0)}(y)\psi_{\bar{3},L}^{(0)}(x) \nonumber \\
&&\quad \quad \quad +\int dy\,f_{\bar{3},L}^{(0)}(y)\bar{\psi}_{\bar{3},L}^{(0)}(x) (i\Gamma^5 \partial_y+h_{\bar{3}})
 f_{\bar{3},L}^{(0)}(y)\psi_{\bar{3},L}^{(0)}(x) \nonumber \\
&& \quad =  i\bar{\psi}_{\bar{3},L}^{(0)}(x) \gamma^\mu \left(\partial_\mu-ig_3 A_{3\mu}^{(0)*}-i{1 \over 3}\sqrt{3 \over 5}g_1 A_{1\mu}^{(0)*} \right)\psi_{\bar{3},L}^{(0)}(x) \nonumber \\
&&\quad \quad \quad + \bar{\psi}_{\bar{3},L}^{(0)}(x) \psi_{\bar{3},L}^{(0)}(x)  \int dy\,f_{\bar{3},L}^{(0)}(y) Q_{\bar{3}} f_{\bar{3},L}^{(0)}(y)\nonumber \\
&&\quad = i\bar{d}^c\gamma^\mu\left(\partial_\mu - i g_3 G_\mu^*-ig_1{1 \over 3}\sqrt{3 \over 5} B^*_\mu \right)d^c\,,
\ee
where we have used (\ref{eq:mode-eq-f}) and (\ref{eq:chirality}).
Repeating the similar calculation for $\Psi_2$, $\Psi_3^\prime$, $\Psi_{3,2}$ and $\Psi_1$,
 we find the resultant effective Lagrangian for ${\cal L}_{\rm f}$ as
\be
 {\cal L}_{\rm f}^{\rm eff}=i\bar{d}^c\gamma^\mu D_\mu d^c + i \bar{l}\gamma^\mu D_\mu l+ i{\rm Tr}[\bar{q}\gamma^\mu D_\mu q]
  +{i \over 2}{\rm Tr}[\bar{U}^c\gamma^\mu D_\mu U^c]+{i \over 2}{\rm Tr}_2[E^+\gamma^\mu D_\mu E^+]\,,
\ee 
 where the covariant derivatives are defined by
\be
D_\mu d^c&=&\partial d^c-ig_3 G_{\mu}^*d^c-ig_1{1 \over 3}\sqrt{3 \over 5}B_{\mu}^*d^c\,, \\
D_\mu l&=&\partial l-ig_2 W_{\mu}^{*}l+ig_1{1 \over 2}\sqrt{3 \over 5}B_{\mu}^{*}l\,, \\
D_\mu q&=&\partial_\mu q+ig_3G_{\mu}q+ig_2q W_{\mu}^{t}+ig_1{1 \over 6}\sqrt{3 \over 5}B_{\mu}q\,,  \\
D_\mu U^c&=&\partial_\mu U^c+2ig_3U^c G_{\mu}^{t}+ig_1{2 \over 3}\sqrt{3 \over 5}B_{\mu}U^c\,, \\
D_\mu E^+ &=&\partial_\mu E^++ig_1\sqrt{3 \over 5}g_1B_{\mu}E^+\,.
\ee
%

%
%
\subsection{Effective Lagrangian of fermion sector 2: SM Yukawa sector}\label{sec:yukawa_eff}
In this subsection, we consider the Yukawa sector with ${\cal H}$.
The Yukawa Lagrangian ${\cal L}_{\rm Yukawa}$ is described by
\begin{eqnarray}
 {\cal L}_{\rm Yukawa}=\Psi_{\bar{5}A}^tCY_{AB}^{(1)}\Psi_{10 B}{\cal H}^\dagger
 +\epsilon_{ijklm}\Psi_{10 A}^{ij~t}Y_{AB}^{(2)}C\Psi_{10 B}^{kl}{\cal H}^m+h.c., \label{yukawa}
\end{eqnarray}
where $C=\gamma_1\gamma_3$ is the charge conjugation matrix in five dimensions, $\epsilon_{ijklm}~(i,j,k,l,m=1,\cdots,5)$ is the five-dimensional 
 Levi-Civita tensor, and $Y_{AB}^{(1)}$ and $Y_{AB}^{(2)}$ are Yukawa couplings.\footnote{
 Since ${\cal H}$ vanishes in the background configuration, ${\cal L}_{\rm Yukawa}$ does not affect the fermion zero modes in the previous subsection. Also, we will find that the ${\cal H}$-Yukawa couplings $Y_{AB}^{(1,2)}$ are smaller than $T$-Yukawa couplings $h_{5,\bar{5},10,\tilde{10}}$, so their contributions can be safely ignored in the previous subsection.}

We derive the effective Lagrangian for the Yukawa sector (\ref{yukawa}).
In particular, we focus on Model 1 of the Higgs sector described in Sec.~\ref{sec:HSWOF}.\footnote{
We expect qualitatively the same results for the Model 2, as the only difference is the Higgs wave functions.}
Substituting (\ref{eq:decomposition_higgs}) and (\ref{eq:comp-f}) into (\ref{yukawa}), we have
\be
{\cal L}_{\rm Yukawa} ={1 \over \sqrt{2}} \Psi_{\bar{3}}^tY^{(1)}\Psi_{3,2}{\cal H}_2^*+{1 \over \sqrt{2} }\Psi_2^t Y^{(1)}\Psi_1{\cal H}_2^*
 +\epsilon_{\alpha\beta\gamma}\Psi_{\bar{3}^\prime}^{\alpha\beta}(Y^{(2)}+Y^{(2)t})\Psi_{3,2}^{\gamma a} \epsilon_{ab}{\cal H}_2^b+h.c., \label{eq:yukawa-dec}
\ee
where $\epsilon_{\alpha\beta\gamma}(\epsilon_{ab})$ is the 3(2)-dimensional Levi-Civita tensor.
The generation indices are suppressed.
We drop the Yukawa couplings including the triplet Higgs since they do not have a normalizable zero mode as shown in Sec.~\ref{sec:HSWOF}.
First we substitute (\ref{eq:he2}) together with (\ref{eq:zh}), (\ref{eq:norma-fac}) and the expansion of the fermions such as 
 (\ref{eq:KKdecomposition}) into (\ref{eq:yukawa-dec}).
Next we only pick up the zero modes of Higgs and the fermions and integrate over $y$.
Then, we obtain the effective Lagrangian of (\ref{yukawa}) as
\be
{\cal L}_{\rm Yukawa}^{\rm eff}=d^c Y_d^{\rm eff}q h_{2, 0}^*+l Y_e^{\rm eff}e^+h_{2,0}^*
 +u^c Y_u^{\rm eff}q h_{2,0}+h.c.\,,
\ee
where 
$Y_u^{\rm eff}$, $Y_d^{\rm eff}$ and $Y_e^{\rm eff}$
are the effective Yukawa couplings given by
\be
Y_u^{\rm eff}&=&2 N_{H,2}
\int dy f_{\bar{3}^\prime,L}^{(0)}(Y^{(2)}+Y^{(2)t})f_{3,2,L}^{(0)}\,, \label{eq:y2} \\
Y_d^{\rm eff}&=&{N_{H,2} \over 2}
 \int dy f_{\bar{3},L}^{(0)}Y^{(1)}f_{3,2,L}^{(0)}\,, \label{eq:y11}\\
Y_e^{\rm eff}&=&{N_{H,2} \over 2}
\int dy f_{2,L}^{(0)}Y^{(1)}f_{1,L}^{(0)}\,, \label{eq:y12}
\ee
where the mode functions such as $f_{\bar{3},L}$ are given in (\ref{eq:zero1}) and (\ref{eq:f2L})--(\ref{eq:f1L}).
We assume that this effective Lagrangian is defined at the GUT scale
$M_X=2.0\times 10^{16}~\rm GeV$ (As for this scale, see, for example, \cite{Hebecker:2002rc}).
Since the normalization constant $N_{H,2}$ can be absorbed into the Yukawa matrices,  the choice of $N_{H,2}$ is merely a convention. 

In the following, we show that the effective Yukawa couplings in (\ref{eq:y2})--(\ref{eq:y12}) can partially realize the four-dimensional Yukawa couplings in the SM,
 which are experimentally determined at the weak scale, by choosing the parameters in the five-dimensional theory.
The strategy is as follows. 
We first run the four-dimensional Yukawa couplings in the SM to the GUT scale $M_X$ from
the weak scale defined at $M_Z=91.2~{\rm GeV}$ by the renormalization group equation (RGE) of the SM as
$Y_{u,d,e}^{\rm SM}(M_Z) \to Y_{u,d,e}^{\rm SM}(M_X)$.
Next we find the parameters, such as $v$, $\Omega$, ${\cal Y}_{2,3}$, $h$ and $\tilde h$, in the five-dimensional theory such that the effective Yukawa couplings $Y_{u,d,e}^{\rm eff}$
 (\ref{eq:y2})-(\ref{eq:y12}) correctly reproduce to the four-dimensional Yukawa coupling $Y^{\rm SM}_{u,d,e}(M_X)$ in the SM at the GUT scale.

However, since dealing with all the Yukawa couplings are a bit complicated, in this paper we simply look at the squared Yukawa couplings   
 $H_a^{\rm SM} =  Y^{\rm SM}_a Y_a^{{\rm SM}\dagger}$ $(a=u, d, e)$ only.
The $H$-RGEs for the squared Yukawa couplings $H_a^{\rm SM}$ at two-loop level are given in \cite{Fusaoka:1998vc}.
In our  analysis, we use the one-loop RGEs which are summarized in \ref{app:C}.
We first take the SM fermion diagonal mass matrices $M_a$ at the renormalization scale
$\mu = M_Z$ \cite{Xing:2011aa} 
\begin{eqnarray}
M_u(M_Z) &=& {\rm diag}(1.38\times 10^{-3},~0.638,~172.1)\,{\rm GeV},\\
M_d(M_Z) &=& {\rm diag}(2.82\times 10^{-3},~57\times 10^{-3},~2.86)\,{\rm GeV},\\
M_e(M_Z) &=& {\rm diag}(0.487\times 10^{-3},~103\times 10^{-3},~1750 \times 10^{-3})\,{\rm GeV}.
\end{eqnarray}
The diagonalized SM Yukawa couplings in the mass basis can be extract as $D^{\rm SM}_a(M_Z) = \sqrt2\, M_a(M_Z)/v_{\rm SM}$ 
where $v_{\rm SM}=246~{\rm GeV}$.
The SM Yukawa couplings $Y^{\rm SM}_a(M_Z)$ in the flavor basis are related to $D_a^{\rm SM}$ via a bi-unitary transformations 
$
D_a = U_L^{a} Y_a U_R^{a\dag}
$
($U_{L,R}^a \in SU(3)$).
As usual, this
leads to the appearance of the Cabibbo-Kobayashi-Maskawa (CKM) matrix 
$V_{\rm CKM} = U_L^{u}U_L^{d\dag}$ 
in the quark charged currents 
 in the mass basis, originating 
from the mismatch of the left-handed rotations in the up- and down-type sectors. 
However, in order to use the $H$-RGE, we need to
redefine the down-type quark fields so as to absorb the CKM matrix in the quark charged currents, 
namely, 
$d_L \to V_{\rm CKM} d_L$,
such that all data for the flavor mixing resides in the down-type Yukawa matrix (the down-sector mixing basis) as,
\begin{eqnarray}
Y_{u,e} \to D_{u,e}  \to D_{u,e}\,,\quad
Y_d \to D_{d} \to V_{\rm CKM}D_d \equiv Y'_d\,.
\end{eqnarray}
Here, we suppressed the superscript SM and the energy scale $(M_Z)$.
As a result, we have the  squared Yukawa couplings $H_a^{\rm SM}$ in the down-sector mixing basis at $\mu=M_Z$:
\begin{eqnarray}
H_u^{\rm SM}(M_Z) &=& D_u^2\,,\qquad D_u = \sqrt{2} v_{\rm SM}^{-1} M_u(M_Z)\,,\\
H_d^{\rm SM}(M_Z) &=& Y'_d Y'_d{}^\dag = V_{\rm CKM}\,D_d^2\,V_{\rm CKM}^\dagger \,,\qquad D_d = \sqrt{2} v_{\rm SM}^{-1} M_d(M_Z)\,,\\
H_e^{\rm SM}(M_Z) &=& D_e^2\,,\qquad D_e = \sqrt{2} v_{\rm SM}^{-1} M_e(M_Z)\,,
\end{eqnarray}
where $V_{\rm CKM}$ is the CKM matrix given by
\begin{eqnarray}
V_{\rm CKM}=
\begin{pmatrix}
 c_{12} c_{13} & s_{12}c_{13} & s_{13}e^{-i\delta} \\
 -s_{12}c_{23}-c_{12}s_{23}e^{i\delta} & c_{12}c_{23}-s_{12}s_{23}s_{13}e^{i\delta} & s_{23}c_{13} \\
 s_{12}s_{23}-c_{12}c_{23}s_{13}e^{i^\delta} &-c_{12}s_{23}-s_{12}c_{23}s_{13}e^{i\delta} & c_{23}c_{13}
\end{pmatrix}
\,.
\end{eqnarray}
Here we denote $s_{ij}\equiv \sin\theta_{ij}$, $c_{ij}\equiv \cos\theta_{ij}$ and $s_{12}=0.22500, s_{13}=0.00369, s_{23}=0.04182$, 
 $\delta=1.144$.
 
\begin{figure}[htbp]
\centering

\begin{subfigure}{0.3\textwidth}
  \centering
  \includegraphics[width=\linewidth]{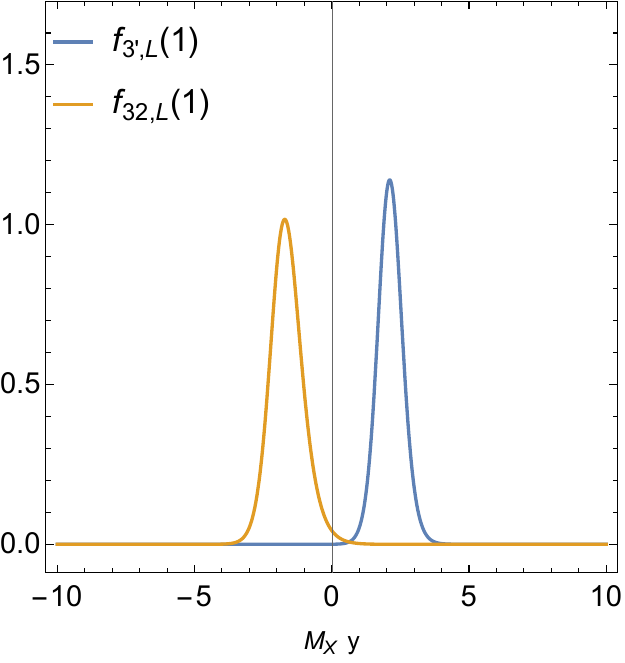}
  \caption{$u$(yellow), $u^c$(blue)}
\end{subfigure}
\hfill
\begin{subfigure}{0.3\textwidth}
  \centering
  \includegraphics[width=\linewidth]{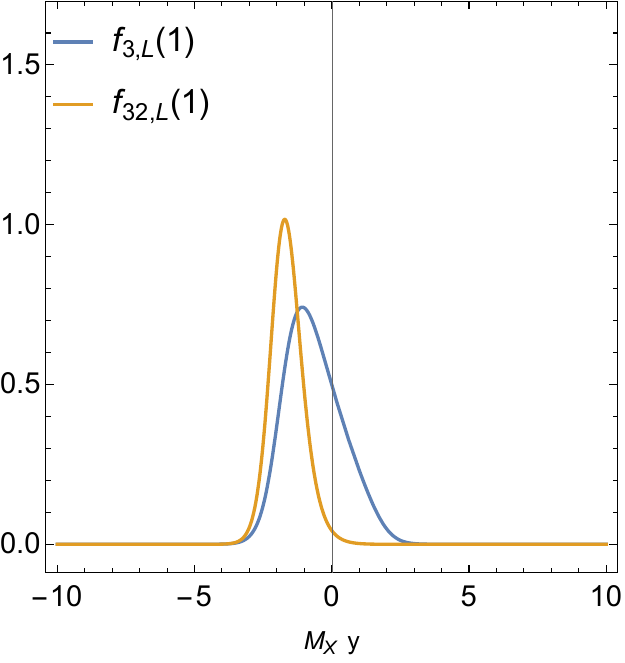}
  \caption{$d$(yellow), $d^c$(blue)}
\end{subfigure}
\hfill
\begin{subfigure}{0.3\textwidth}
  \centering
  \includegraphics[width=\linewidth]{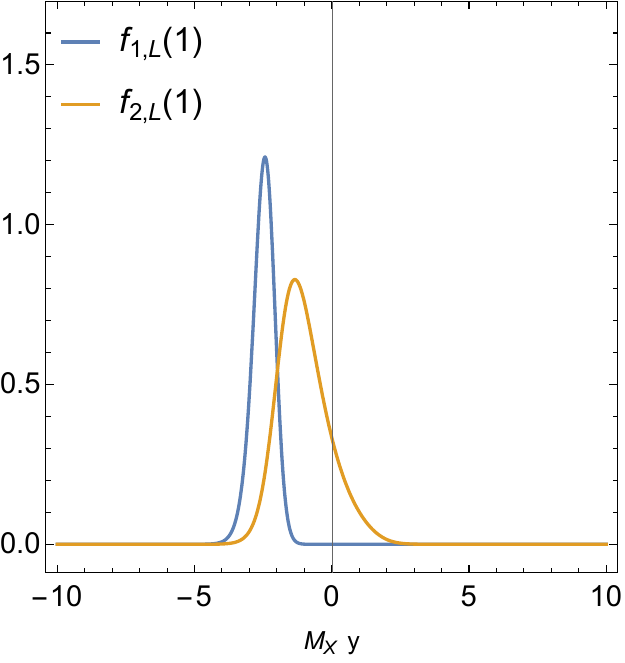}
  \caption{$l_1$(yellow), $e^+$(blue)}
\end{subfigure}

\vspace{0.5em}

\begin{subfigure}{0.3\textwidth}
  \centering
  \includegraphics[width=\linewidth]{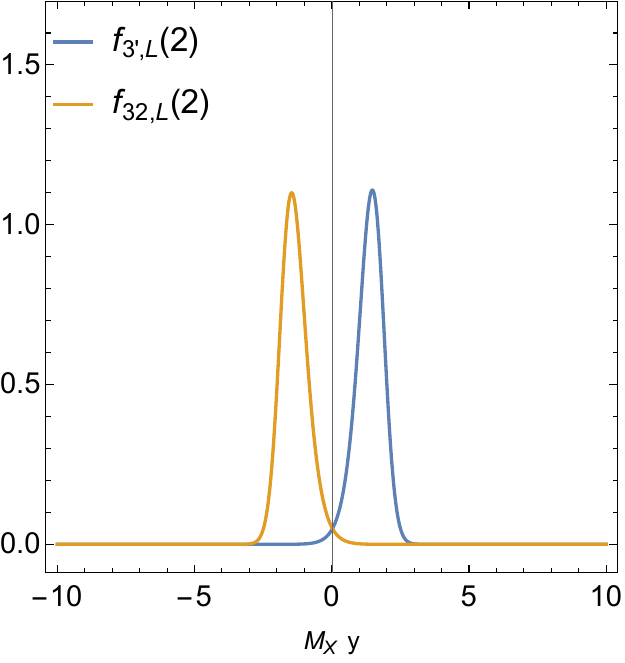}
  \caption{$c$(yellow), $c^c$(blue)}
\end{subfigure}
\hfill
\begin{subfigure}{0.3\textwidth}
  \centering
  \includegraphics[width=\linewidth]{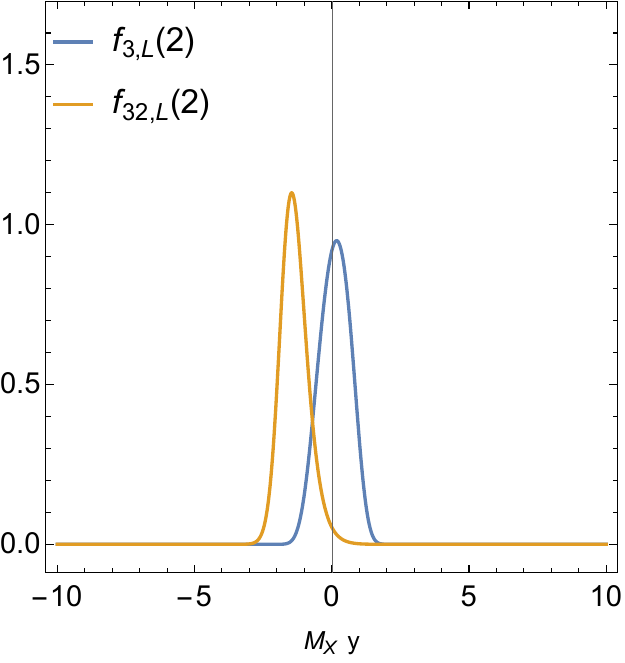}
  \caption{$s$(yellow), $s^c$(blue)}
\end{subfigure}
\hfill
\begin{subfigure}{0.3\textwidth}
  \centering
  \includegraphics[width=\linewidth]{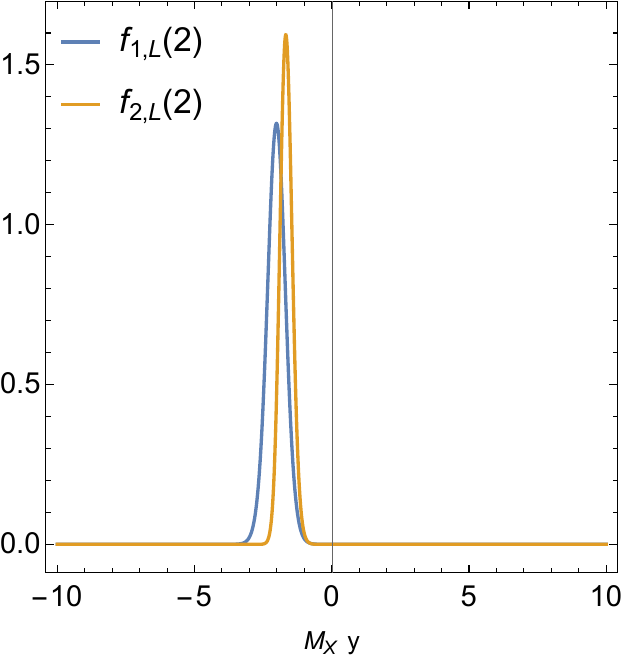}
  \caption{$l_2$(yellow), $\mu^+$(blue)}
\end{subfigure}

\vspace{0.5em}

\begin{subfigure}{0.3\textwidth}
  \centering
  \includegraphics[width=\linewidth]{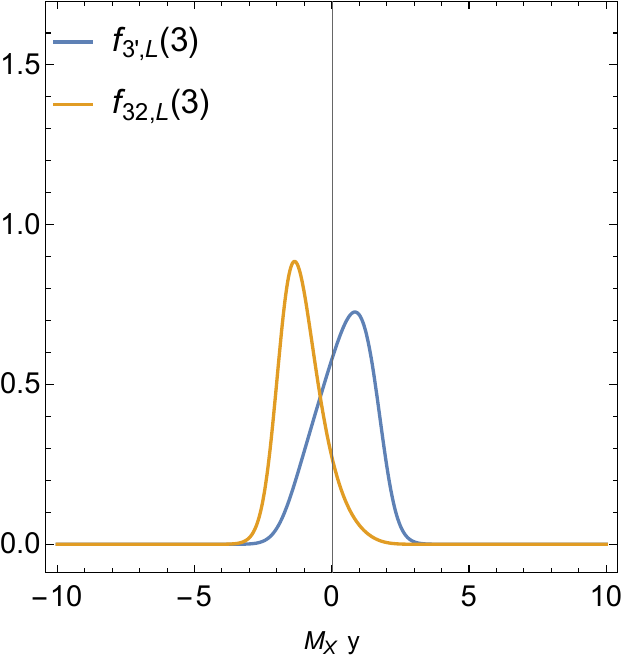}
  \caption{$t$(yellow), $t^c$(blue)}
\end{subfigure}
\hfill
\begin{subfigure}{0.3\textwidth}
  \centering
  \includegraphics[width=\linewidth]{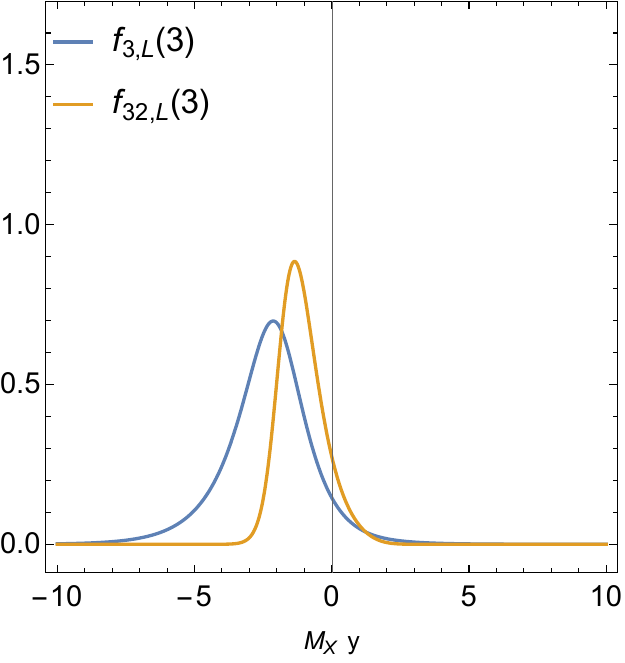}
  \caption{$b$(yellow), $b^c$(blue)}
\end{subfigure}
\hfill
\begin{subfigure}{0.3\textwidth}
  \centering
  \includegraphics[width=\linewidth]{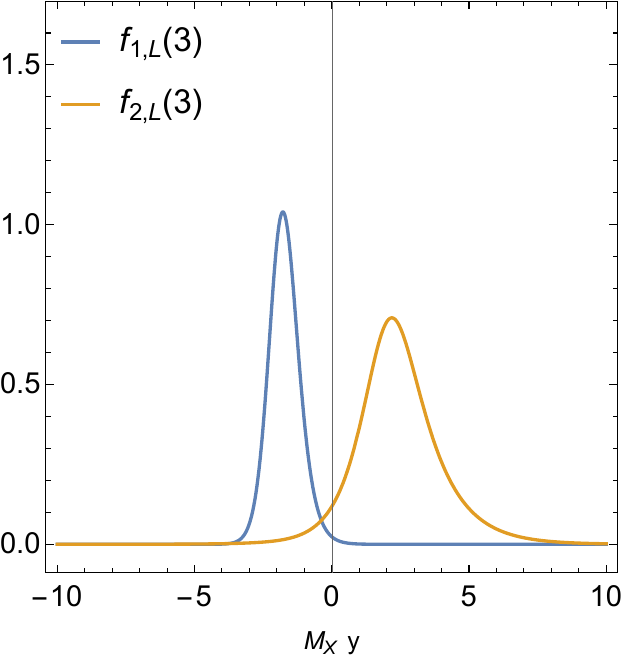}
  \caption{$l_3$(yellow), $\tau^+$(blue)}
\end{subfigure}

\caption{Nine overlaps of the normalized fermion zero mode functions.}
\label{fig:3x3}
\end{figure}

Now, we run  $H_a^{\rm SM}$ at $M_Z$ to the  GUT scale at $M_X$.
Note that $H_a(M_X)$ is not diagonalized in general, but we can diagonalize it again by using $U_L^a$ only as
$H_a \to (U_L^{a} Y_a U_R^a) (U_L^{a} Y_a U_R^a)^\dag = U_L^{a} H_a U_L^{a\dag}$.
%
%
%
%
%
%
%
The resultant diagonal matrices $\sqrt{H_a}(M_X)$ are found as
\begin{eqnarray}
\sqrt{H_u}(M_X)\big|_{\rm diag}&=&{\rm diag}(3.24\times 10^{-6}, 1.50\times 10^{-3}, 4.61\times 10^{-1})\,, \label{eq:RGEYu} \\
\sqrt{H_d}(M_X)\big|_{\rm diag}&=& {\rm diag}(6.84\times 10^{-6}, 1.38\times 10^{-4}, 6.08\times 10^{-3})\,,  \label{eq:RGEYd} \\
\sqrt{H_e}(M_X)\big|_{\rm diag}&=&{\rm diag}(2.75\times 10^{-6}, 5.80\times 10^{-4}, 9.86\times 10^{-3})\,.\label{eq:RGEYe}
\end{eqnarray}
Let us compare these originated from the SM with $\sqrt{H_a^{\rm eff}} = \sqrt{Y_a^{\rm eff}Y_a^{\rm eff \dag}}$ with the effective Yukawa couplings $
 Y_a^{\rm eff}$ in (\ref{eq:y11}), (\ref{eq:y12}), (\ref{eq:y2}) originated by the five dimensional theory.
Since our focus is solely on reproducing the fermion mass spectrum, we compare the eigenvalues of $\sqrt{H}$.
We find that the following set of parameters in the five-dimensional Lagrangian can reproduce good agreement:
\begin{eqnarray}
v=M_X^{\frac{3}{2}}\,,\quad
\Omega=M_X,\quad 
{\cal Y}_2=-{\cal Y}_3=2M_X^{-1}\,,\quad
c = 0.13 M_X^{\frac{3}{2}}\,,
\end{eqnarray}
where ${\cal Y}_2={\cal Y}_3=2M_X^{-1}$ means the separation of the walls is $d = 4 M_X^{-1}$ as shown in Fig.~\ref{domain2}.
The parameter $c$ is in the $\beta_H(\hat T)$ given  in (\ref{eq:beta_H}) and it gives $N_{H,2}=0.2 M_X^{1/2}$ by (\ref{eq:norma-fac}).
We also choose the 5 dimensional Yukawa couplings for $\hat T$ and $T_0$ in the following way
\begin{eqnarray}
h_{5,1}=-0.227,~\tilde{h}_{5,1}=0.800,~h_{5,2}=-8.00,~\tilde{h}_{5,2}=6.00,~h_{5,3}=1.18,~\tilde{h}_{5,3}=0.170\,, \\
h_{10,1}=-9.06,~\tilde{h}_{10,1}=1.00,~h_{10,2}=-6.6,~\tilde{h}_{10,2}=1.91,~h_{10,3}=-1.85,~\tilde{h}_{10,3}=1.00\,,
\end{eqnarray}
together with the 5-dimensional Yukawa couplings for ${\cal H}$ in the mass basis as
\begin{eqnarray}
Y^{(1)} = {\rm diag}(x_1,~x_2,~x_3)\,,\quad
Y^{(2)} = {\rm diag}(y_1,~y_2,~y_3)\,.
\end{eqnarray}
These are all in the unit of $M_X^{-\frac{1}{2}}$. The fermion zero mode functions are shown in Fig.~\ref{fig:3x3}.
By using Eqs.~(\ref{eq:y2})--(\ref{eq:y12}), we find
\begin{eqnarray}
\sqrt{H_u^{\rm eff}} &=& {\rm diag}\left(|y_1| \times 1.89 \times 10^{-4}\,, |y_2| \times 3.55 \times 10^{-3}\,, |y_3| \times 3.67 \times 10^{-1}\right)\,,\\
\sqrt{H_d^{\rm eff}} &=& {\rm diag}\left(|x_1| \times 7.31\times 10^{-2}\,, |x_2| \times 1.55 \times 10^{-2}\,, |x_3| \times 7.25 \times 10^{-2}\right)\,,\\
\sqrt{H_e^{\rm eff}} &=& {\rm diag}\left(|x_1| \times 2.74 \times 10^{-6}\,, |x_2| \times 5.80 \times 10^{-4}\,, |x_3| \times 2.36 \times 10^{-4}\right)\,.
\end{eqnarray}
We find $|x_a|$ and $|y_a|$ that provides closest match between $\sqrt{H_a^{\rm SM}}(M_X)$ and  $\sqrt{H_u^{\rm eff}}$ up to three decimals. It is not trivial because we have overdetermined system of 9 equations for 6 variables $|x_a|$ and $|y_a|$. 
Choosing the 
6
parameters in the following way,
\begin{eqnarray}
(|x_1|,~|x_2|,~|x_3|) &=& (9.35\times 10^{-5}, 8.91\times 10^{-3}, 8.39\times 10^{-2})\,, \\
(|y_1|,~|y_2|,~|y_3|) &=& (1.72 \times 10^{-2}, 4.21\times 10^{-1}, 1.25)\,,
\end{eqnarray}
the resultant numerical matrices are found to be
\begin{eqnarray}
\sqrt{H^{\rm eff}_u}&=&{\rm diag}(3.24\times 10^{-6}, 1.50\times 10^{-3}, 4.61\times 10^{-1})\,, \\
\sqrt{H^{\rm eff}_d}&=& {\rm diag}(6.84\times 10^{-6}, 1.38\times 10^{-4}, 6.08\times 10^{-3})\,, \\
\sqrt{H^{\rm eff}_e}&=&{\rm diag}(2.75\times 10^{-6}, 5.80\times 10^{-4}, 9.86\times 10^{-3})\,,
\end{eqnarray}
which are in perfect agreement with the RGE-evolved SM Yukawa matrices in Eqs.~(\ref{eq:RGEYu})-(\ref{eq:RGEYe})
within the displayed numerical precision.

Importantly, in minimal four-dimensional $SU(5)$ GUTs, the unified Yukawa structure generically predicts the relation $Y_d = Y_e^{T}$ at the unification scale, leading to phenomenologically unrealistic fermion mass relations. 
Conventional resolutions typically involve extending the Higgs sector, for example by introducing a $\mathbf{45}$ representation as in the Georgi--Jarlskog mechanism~\cite{Georgi:1979df}, or by invoking higher-dimensional operators that modify the Yukawa structure~\cite{Ellis:1979fg}.

In contrast, our framework avoids this problem without introducing such ingredients. 
The model employs only the adjoint $\mathbf{24}$ representation for $SU(5)$ breaking, as in minimal $SU(5)$, and does not rely on higher-dimensional operators. 
Instead, the theory is formulated in five dimensions, where the effective four-dimensional Yukawa matrices arise from overlap integrals of localized fermion zero-mode 
functions in the extra dimension. 
Although the underlying five-dimensional Yukawa couplings respect $SU(5)$ unification, the generation-dependent wave-function overlaps induce non-universal corrections to the effective four-dimensional Yukawa matrices. 
As a result, the usual $SU(5)$ fermion mass relations are relaxed, allowing for a realistic pattern of fermion masses and mixings.

\subsection{Effective Lagrangian for the Higgs sector}\label{sec:Higgs_eff}

Finally, we shall discuss the electroweak symmetry breaking in the Higgs sector in the effective Lagrangian.

\subsubsection{Model 1}\label{sec:Model_1}

We shall modify the five-dimensional Lagrangian for the model, (\ref{eq:beta_Higgs}).
We first note that the final result in this section is valid for any $\beta_H$.
We propose a starting Lagrangian by adding a scalar potential as a perturbation
 which will be a seed of the electroweak symmetry breaking at a low energy:
\be
{\cal L}_{\rm H} = \tr \left[\beta_H(\hat T, T_0)^2\, \left\{({\cal D}_M {\cal H})^\dagger {\cal D}^M{\cal H} -V_{\rm pert}\right\}\right]\,,
\ee
where
\be
V_{\rm pert}=-\tilde{\mu}^2{\cal H}^\dagger {\cal H}+\tilde{\lambda}({\cal H}^\dagger{\cal H})^2\,. \label{eq:Higgs_pert_pot}
\ee
We assume that this potential becomes relevant at energies which are much lower than the GUT scale.
Therefore $V_{\rm pert}$ does not affect the results concerning the double-triplet splitting problem, which are mentioned above.
Decomposing the Higgs as (\ref{eq:decomposition_higgs}) and dropping the triplet part ${\cal H}_3$, we have the effective
Lagrangian for the standard model Higgs ${\cal H}_2$:
\be
{\cal L}_{\rm H}=\beta_{H,2}^2\left\{(D_\mu{\cal H}_2)^\dagger D^\mu{\cal H}_2-V^\prime \right\}\,, \label{eq:mwf2}
\ee
where $D_\mu {\cal H}_2=\partial_\mu {\cal H}_2-i{\cal A}_{2\mu}{\cal H}_2+i{1 \over 2}\sqrt{3 \over 5}{\cal A}_{1\mu}{\cal H}_2$ and
\be
V^\prime=\beta_{H,2}^2\left\{(\partial_y{\cal H}_2)^\dagger \partial_y{\cal H}_2-\tilde{\mu}^2{\cal H}_2^\dagger {\cal H}_2+\tilde{\lambda}({\cal H}_2^\dagger{\cal H}_2)^2\right\}\,.
\ee
From (\ref{eq:cn}) and (\ref{eq:he2}) with (\ref{eq:zh}), we can write
\be
{\cal H}_2=N_{H,2}h_{2,0}+{\rm massive~modes}\,. 
\label{eq:h2-zero}
\ee
Here $N_{H,2} = 1/\sqrt{\int dy~\beta_{H,2}^2}$ as is given in Eq.~(\ref{eq:norma-fac}).
Substituting (\ref{eq:h2-zero}) into (\ref{eq:mwf2}) and integrating over $y$, we obtain
\be
{\cal L}_{\rm H}^{\rm eff}=(D_\mu h_{2,0})^\dagger D^\mu h_{2,0}-V^{\prime \rm eff}\,,
\ee
where $D_\mu h_{2,0}=\partial_\mu h_{2,0}-ig_2 W_{\mu}h_{2,0}+ig_1{1 \over 2}\sqrt{3 \over 5}B_{\mu}h_{2,0}$ and
\be
V^{\prime \rm eff}=-\tilde{\mu}^2h_{2,0}^\dagger h_{2,0}+\tilde{\lambda}^\prime(h_{2,0}^\dagger h_{2,0})^2\,, \label{eq:Higgs_eff_pot}
 \ee
with $\tilde{\lambda}^\prime=N_{H,2}^2\tilde{\lambda}$.

\subsubsection{Model 2}
We start with the five-dimensional Lagrangian (\ref{eq:H_lag}).
In the effective Lagrangian the colored Higgs ${\cal H}_3$ is simply dropped since it is so heavy.
For the standard model Higgs ${\cal H}_2$, we only consider the lowest mode. 
Then, the Lagrangian (\ref{eq:H_lag}) becomes
\be
{\cal L}_{\rm H}=(D_M{\cal H}_2)^\dagger D^M{\cal H}_2
 -{\cal H}_2^\dagger U_2 {\cal H}_2
 -\lambda_{\rm H}({\cal H}_2^\dagger{\cal H}_2)^2\,, \label{eq:ef}
\ee
where $D_M{\cal H}_2=\partial_M{\cal H}_2-i{\cal A}_{2M}{\cal H}_2+i{1 \over 2}\sqrt{3 \over 5}{\cal A}_{1M}{\cal H}_2$.
Note that we do not need to add additional potential terms because the starting Lagrangian (\ref{eq:H_lag}) already has a seed of the electroweak symmetry breaking.
We derive the effective Lagrangian in the $\tilde{d}\rightarrow 0$ limit, so that we can utilize the analytic form of 
 the mode functions (\ref{eq:af}).
Substituting (\ref{eq:KKdecomposition}) and (\ref{eq:or}), integrating over $y$, and picking up the zero mode $h_{2,0}$ in Eq.~(\ref{eq:KKdecomposition}), 
 we obtain
\be
{\cal L}_{\rm H}^{\rm eff}=(D_\mu h_{2,0})^\dagger D^\mu h_{2,0}-V_{h}\,,
\ee
where $D_\mu h_{2,0}=\partial_\mu h_{2,0}-ig_2 W_\mu h_{2,0}+ig_1{1 \over 2}\sqrt{3 \over 5}B_{\mu}h_{2,0}$ and
\be
 V_{h}&=&\int dy\left\{(h_{2,0}^\dagger h_{2,0})(\partial_y u_0(y))^2 +h_{2,0}^\dagger U_0 h_{2,0} (u_0(y))^2+\lambda_{\rm H}(h_{2,0}^\dagger h_{2,0})^2 (u_0(y))^4\right\}\\
 &=&\Omega_{\rm H}^2h_{2,0}^\dagger h_{2,0}+V_{\rm K}+V_{\rm P}+V_4\,,\\
V_{\rm K}&=&h_{2,0}^\dagger h_{2,0} \int dy (\partial_y u_0(y))^2\,, \label{eq:VK}\\
V_{\rm P}&=&h_{2,0}^\dagger h_{2,0} \int dy (u_0(y))^2{-U_0 \over \cosh\Omega\left(y-{{\cal Y}_2-{\cal Y}_3 \over 2}\right)\cosh\Omega\left(y+{{\cal Y}_2-{\cal Y}_3 \over 2}\right)}\,, \label{eq:VP}\\
V_4&=&\lambda_{\rm H}(h_{2,0}^\dagger h_{2,0})^2\int dy (u_0(y))^4\,. \label{eq:V4}
\ee
Note that ${\cal A}_{\alpha 5}$ is simply dropped in the covariant derivative since their zero modes do not exist.
We can analytically evaluate (\ref{eq:VK}), (\ref{eq:VP}) and (\ref{eq:V4}) with the use of (\ref{eq:zero-H}) as
\be
V_{\rm K}&=&{s^2 \over 2s+1}\Omega^2 h_{2,0}^\dagger h_{2,0}\,, \\
V_{\rm P}&=&-U_0{2s \over 2s+1} h_{2,0}^\dagger h_{2,0}\,, \\
V_4&=&\lambda_{\rm H}\Omega {B\left({1\over 2},2s\right) \over \left(B\left({1 \over 2},s\right)\right)^2}(h_{2,0}^\dagger h_{2,0})^2\,,
\ee
where $B(p,q)$ is the beta function. 
Collecting the above results, we have the effective potential as
\be
 V_h=\mu^2h_{2,0}^\dagger h_{2,0}+{\lambda \over 2}(h_{2,0}^\dagger h_{2,0})^2\,, 
\ee
with
\be
 \mu^2&=&{s^2 \over 2s+1}\Omega^2-{2s \over 2s+1}U_0+\Omega_{\rm H}^2\,,\\
 {\lambda \over 2}&=&\lambda_{\rm H}^2\Omega{B\left({1\over 2},2s\right) \over \left(B\left({1 \over 2},s\right)\right)^2}\,.
\ee
%

%
%
\section{Conclusion}\label{sec:conc}

\begin{figure}[htbp]
\centering
\includegraphics[width=17cm]{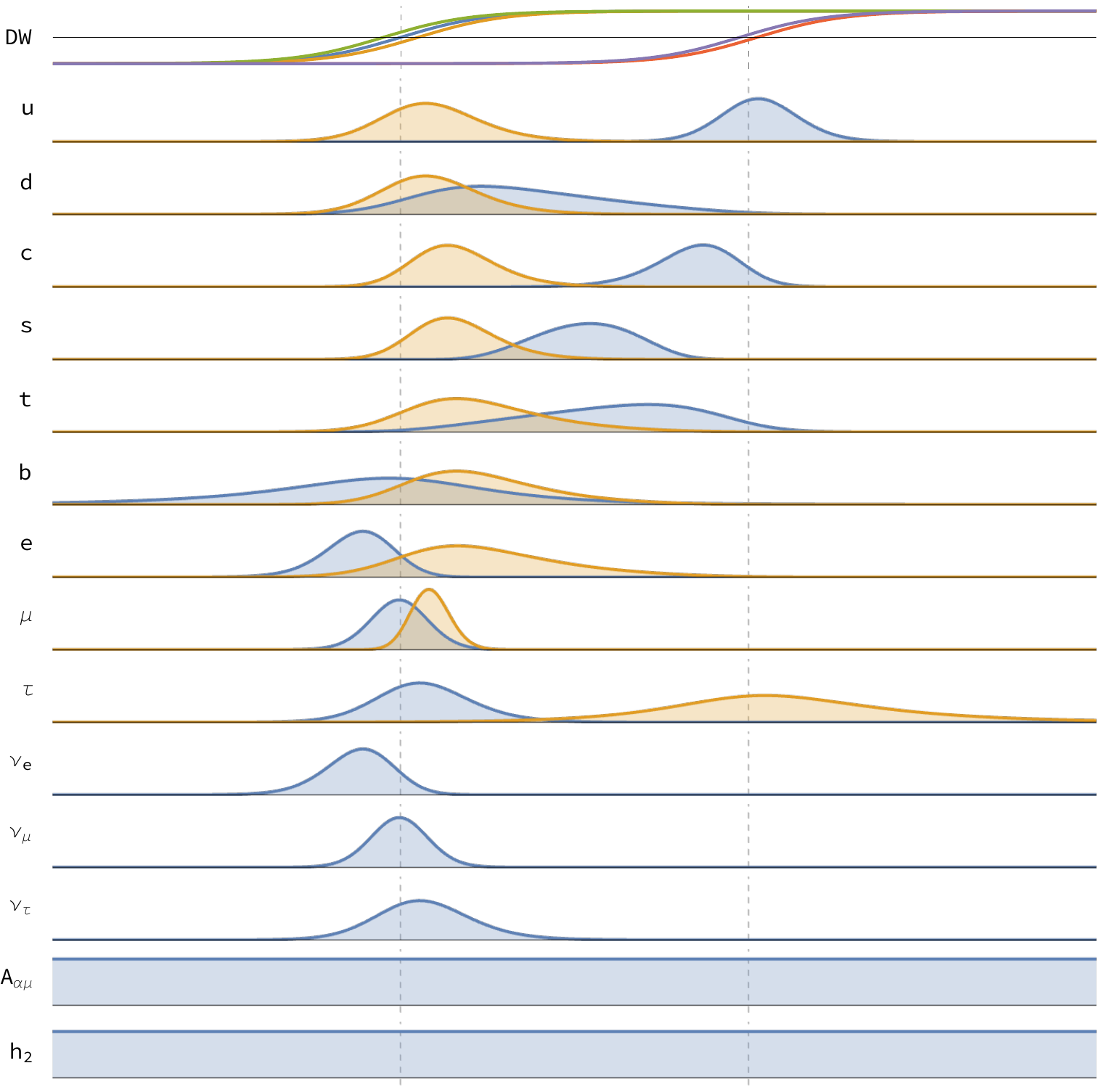}
\caption{The summary of our $SU(5)$ GUT in the five dimensions. The top line corresponds to the 3-2 splitting domain walls, followed by the six quarks and six leptons. The last two rows show the flat zero mode functions for the gauge fields and the SM Higgs doublet in the model 1, respectively.}
\label{fig:summary}
\end{figure}

In this work, we have constructed a five-dimensional $SU(5)$ grand unified theory formulated on domain walls, 
 in which both gauge and matter fields are dynamically localized. 
The framework naturally realizes a brane-world scenario without assuming the existence of branes by hand, as the 
 four-dimensional universe emerges on a domain wall configuration generated by spontaneous symmetry breaking in the bulk. 
The overall structure of the present framework is summarized schematically in Fig.~\ref{fig:summary}.
The top panel of Fig.~\ref{fig:summary} illustrates the energetically preferred 3-2 split domain-wall configuration, 
 which spontaneously breaks $SU(5)$ down to the SM gauge group $SU(3)\times SU(2) \times U(1)$.
This configuration stabilizes by the inclusion of an additional potential term, leading to a robust domain-wall background 
 suitable for localizing the SM fields as discussed in Sec.~\ref{sec:moduli_stabilization}.

The middle panels of Fig.~\ref{fig:summary} summarize the localization of the quark and lepton zero modes. 
Their localized wave-function profiles provide the origin of the hierarchical Yukawa couplings in the four-dimensional effective theory.
 
In Sec.~\ref{sec:gauge_eff}, the localization of gauge fields is achieved by introducing a field-dependent gauge kinetic term, effectively realizing a semiclassical analog of the confinement in the bulk. 
This mechanism ensures that massless gauge bosons are localized to the vicinity of the domain wall, independent of the detailed form of the field-dependent coupling. We should emphasize that the localization of the gauge field is accomplished in terms of ``dressed gauge field'' $A_{\alpha\mu}(x^\mu,y) \simeq g_\alpha \beta_\alpha(y) A_{\alpha\mu}^{(0)}(x^\mu) + \cdots$ with the zero mode function $\beta_\alpha(y)$ which can be any square integrable function. This leads to the fact that the zero mode function for the original gauge field is flat in the $y$ coordinate as ${\cal A}_{\alpha\mu}(x^\mu,y) = \beta_\alpha(y)^{-1}A_{\alpha\mu}(x^\mu,y) \simeq g_\alpha A_{\alpha\mu}^{(0)}(x^\mu) + \cdots$.
The flatness is physically important because it ensures the universality of the gauge coupling constants $g_\alpha$ for the non-Abelian gauge symmetries. The flat zero mode function is shown in the row at the second from the bottom of Fig.~\ref{fig:summary}.


The most distinctive feature of our construction is the treatment of the Higgs sector as discussed in Sec.~\ref{sec:Higgs}, 
 which resolves the long-standing doublet-triplet splitting problem of $SU(5)$ GUTs. 
We have proposed two distinct models. 
In Model 1, the doublet-triplet splitting is achieved dynamically through a non-minimal, topological coupling between 
  the Higgs field and the domain-wall background.
The electroweak Higgs doublet possesses a normalizable zero mode $h_{2,0}(x^\mu)$ localized between the separated walls, 
 while the colored triplet mode $h_{3,0}(x^\mu)$ remains non-normalizable and hence decoupled as shown in Fig.~\ref{fig:bega_H}.
Similarly to the localization of the gauge fields, the localized zero mode function for $h_{2,0}(x^\mu)$ appears 
 in the ``dressed Higgs field'' as $H_I(x^\mu,y) = N_{H,I} \beta_{H,I}(y) h_{I,0}(x^\mu) + \cdots$. 
Therefore, the zero mode function of the original Higgs field is flat as ${\cal H}_I(x^\mu,y) = \beta_{H,I}(y)^{-1} H_I(x^\mu,y) = N_{H,I} h_{I,0}(x^\mu) + \cdots$. 
The flatness of the Higgs doublet zero mode function is not necessary but it makes the calculation of the effective Yukawa 
 couplings quite simple as we seen in Sec.~\ref{sec:yukawa_eff}.
The flat zero mode function is shown at the bottom row of Fig.~\ref{fig:summary}.
Note that the massless doublet appears as a topologically protected edge state, analogous to the Jackiw-Rebbi fermion, and no fine-tuning of 
 parameters is required.
In Model~2, a potential coupling between the Higgs and the adjoint scalar fields is introduced. 
The resulting potential repels the colored Higgs triplet from the domain wall, so that it possesses neither localized nor light modes. 
Consequently, only the Higgs doublet needs to be examined for possible light or tachyonic modes. 
The correct pattern of electroweak symmetry breaking can then be realized by an appropriate fine-tuning of the model parameters.

We have also derived the four-dimensional effective theory by integrating out the extra dimension. 
The effective Lagrangian successfully reproduces the SM structure at low energies, including the correct 
 gauge kinetic terms, fermion sectors, and Yukawa interactions. 
Furthermore, renormalization group analysis demonstrates that the SM Yukawa couplings at the weak scale 
 can be reproduced from appropriate choices of the five-dimensional parameters, establishing the phenomenological viability 
 of the framework.

Although the present work has focused on the theoretical construction of a realistic dynamical 
 $SU(5)$ brane-world scenario, the completion of the Higgs sector and Yukawa interactions considerably broadens its  
 phenomenological scope. 
In particular, it will be interesting to revisit proton decay by extending the analysis of Ref.~\cite{Arai:2017lfv} to 
 include the Higgs sector and Yukawa interactions developed in this work. 
The present framework also provides a realistic $SU(5)$ setting for investigating phenomenological aspects of 
 dynamical brane-world models, including collider signatures of Kaluza--Klein gauge bosons and fermions, and Higgs  
 phenomenology and flavor observables associated with non-universal Kaluza--Klein gauge 
 couplings~\cite{Okada:2017omx, Okada:2018von, Okada:2019fgm}. 
These issues are left for future work.
%
%
\appendix
\def\thesection{Appendix \Alph{section}}
\section{The effective potential for domain walls}
\label{app:A}
\renewcommand{\theequation}{\Alph{section}.\arabic{equation}}

In this appendix, we discuss the calculation of  the effective potential of Sec.~\ref{sec:model}. 

The problem at hand is to calculate the integral
\begin{equation}
\int\limits_{-\infty}^{\infty}\diff x\,V_1\bigl(T_{\rm wals}, \tr \bigl(T_{\rm walls})\bigr) = V_{\rm eff}\bigl(R_1, \ldots, R_5\bigr)\,,
\end{equation}
where we denoted that $V_1$ can be function of both $T_{\rm walls}$ and its trace. 
In general, such integral can be broken down to evaluation of many pieces of the form
\begin{equation}
\int\limits_{-\infty}^{\infty}\diff x\, \tanh^{m_i}(x-R_{i}) \tanh^{m_j}(x-R_{j}) \ldots  \tanh^{m_k}(x-R_{k})\,,
\end{equation}
where $m_i$'s are some integers. All of these integrals can be related (by differentiation with respect to $R_i$'s or by taking limits) to some combination of the following \emph{generating functions}:
\begin{align}
G_1(i) \equiv & \int\limits_{-\infty}^{\infty}\diff x\bigl(\tanh(x-R_i)-\tanh(x)\bigr)\,, \\
G_2(i, j) \equiv & \int\limits_{-\infty}^{\infty}\diff x\bigl(\tanh(x-R_i)\tanh(x-R_j)-1\bigr)\,, \\
G_3(i, j,k) \equiv & \int\limits_{-\infty}^{\infty}\diff x\bigl(\tanh(x-R_i)\tanh(x-R_j)\tanh(x-R_k)-\tanh(x)\bigr)\,, \\
G_4(i, j,k,l) \equiv & \int\limits_{-\infty}^{\infty}\diff x\bigl(\tanh(x-R_i)\tanh(x-R_j)\tanh(x-R_k)\tanh(x-R_l)-1\bigr)\,, \\
\vdots & \nonumber  
\end{align} 
Notice that for odd number of $R_i$'s we subtracted $\tanh(x)$, while for even number we subtracted 1 to have manifest convergence.

Using the identity
\begin{equation}
\tanh(x-R_i)\tanh(x-R_j)-1 = \frac{\tanh(x-R_i)-\tanh(x-R_j)}{\tanh(R_i-R_j)}\,,
\end{equation}
it is possible to establish  recurrence relations:
\begin{equation}\label{eq:reccur}
G_n(i, j, k, \ldots, l) = \frac{G_{n-1}(i, k,\ldots, l)-G_{n-1}(j, k,\ldots, l)}{\tanh\bigl(R_i-R_j\bigr)}+G_{n-2}(k,\ldots, l)\,,
\end{equation}
that, in principle, allow for the computation of any given $G_n$ starting from $G_1(i) = -2 R_i$.

After a bit of algebra we can obtain explicit forms of a first few generating functions as:
\begin{align}
G_2(i, j) = & -2\Bigl( \frac{R_i}{\tanh_{i-j}}+ \frac{R_j}{\tanh_{j-i}}\Bigr)\,, \\
G_3(i,j,k) = & -2\Bigl( \frac{R_i}{\tanh_{i-j}\tanh_{i-k}}+ \frac{R_j}{\tanh_{j-i}\tanh_{j-k}}+\frac{R_k}{\tanh_{k-i}\tanh_{k-j}}\Bigr)\,, \\
G_4(i,j,k,l) = & -2\Bigl( \frac{R_i}{\tanh_{i-j}\tanh_{i-k}\tanh_{i-l}}+ \frac{R_j}{\tanh_{j-i}\tanh_{j-k}\tanh_{j-l}}\nonumber \\ &+\frac{R_k}{\tanh_{k-i}\tanh_{k-j}\tanh_{k-l}}+\frac{R_l}{\tanh_{l-i}\tanh_{l-j}\tanh_{l-k}}\Bigr)\,,
\end{align}
where we used a shorthand $\tanh_{i-j} \equiv \tanh(R_i - R_j)$.
From this, one can make a general guess
\begin{equation}
G_n(i_1 i_2 \ldots i_n) = -2 \sum\limits_{j = i_1}^{i_n}\frac{R_j}{\prod\limits_{\begin{smallmatrix}k=i_1 \\ k \not = j\end{smallmatrix}}^{i_n}\tanh\bigl(R_{j}-R_{k}\bigr)}\,.
\end{equation}
This expression can be then used to check the validity of the recurrence \refer{eq:reccur}.

Note that all $G_n$'s are totally symmetric functions of their arguments. This fact is useful when summing over all positions (which occurs due to presence of traces of $T_{\rm wall}$'s). To be concrete, the following identities (that holds for any symmetric function) are useful when evaluating the effective potential:
\begin{align}
\sum\limits_{i, j} G_2(i,j) = & 2\sum\limits_{i> j} G_2(i,j) +\sum\limits_{i}G_2(i,i)\,, \\
\sum\limits_{i, j,k} G_3(i,j,k) = & 6\sum\limits_{i> j>k} G_3(i,j,k) +3\sum\limits_{i>j}\Bigl(G_3(i,j,j)+G_3(i,i,j)\Bigr)+\sum\limits_{i}G_3(i,i,i)\,, \\
\sum\limits_{i, j,k,l} G_4(i,j,k,l) = & 24\sum\limits_{i> j>k>l} G_4(i,j,k,l) +12\sum\limits_{i>j>k}\Bigl(G_4(i,j,k,k)+G_4(i,j,j,k)+G_4(i,i,j,k)\Bigr)\nonumber \\ &
+\sum\limits_{i>j}\Bigl(6G_4(i,i,j,j)+4G_4(i,j,j,j)+4G_4(i,i,i,j)\Bigr)+\sum\limits_{i}G_4(i,i,i,i)\,. 
\end{align}

Let us illustrate our methods on evaluating 
\begin{equation}
V_{\rm eff} = v^2 \lambda \int\limits_{-\infty}^{\infty}\diff y\, \Bigl(\alpha \Tr \bigl(\hat T_{\rm wall}^4\bigr)-v^2 \mu^2 \Tr\bigl(\hat T_{\rm wall}^2\bigr)\Bigr)\,. 
\end{equation}
This integral only converges when the outside vacua belong to the $\langle 0 \rangle$ set. Hence, we can set $s_1 = \ldots s_ 5 =1$ without loss of generality. In order to simplify our task, let us evaluate each term separately. The second term gives
\begin{align}
-\lambda v^2 \mu^2 \int\limits_{-\infty}^{\infty}\diff y\,  \Tr\bigl(\hat T_{\rm wall}^2\bigr) = & \frac{\sqrt{\lambda}\, v^3 \mu^2}{5} \sum\limits_{i,j}\int\limits_{-\infty}^{\infty}\diff x\, \Bigl(\tanh(x-R_i)\tanh(x-R_j)-\tanh^2(x-R_i)\Bigr) \nonumber \\
= &  \frac{\sqrt{\lambda}\, v^3 \mu^2}{5}\sum_{i,j}\Bigl(G_2(i,j)-G_2(i,i)\Bigr) \nonumber \\
= &   \sqrt{\lambda}\, v^3 \mu^2\Bigl(8-\frac{4}{5}\sum\limits_{i>j}\frac{R_i-R_j}{\tanh\bigl(R_i-R_j\bigr)}\Bigr)\,.
\end{align}
This piece of the effective potential describes a repulsion for any pair of walls.

The second piece reads:
\begin{gather}
\lambda \alpha \int\limits_{-\infty}^{\infty}\diff y\,  \Tr\bigl(\hat T_{\rm wall}^4\bigr) =  \frac{\sqrt{\lambda}\, v^3 \alpha}{125} \sum\limits_{i,j,k,l}\int\limits_{-\infty}^{\infty}\diff x\, \Bigl(-3\tanh(x-R_i)\tanh(x-R_j)\tanh(x-R_k)\tanh(x-R_l)\nonumber \\
+2\bigl(\tanh^2(x-R_i)\tanh(x-R_j)\tanh(x-R_k) +\tanh(x-R_i)\tanh^2(x-R_j)\tanh(x-R_k)\nonumber \\ 
+\tanh(x-R_i)\tanh(x-R_j)\tanh^2(x-R_k)\bigr)-4 \tanh^3(x-R_i)\tanh(x-R_j)+\tanh^4(x-R_i)\Bigr) \nonumber \\
=   \frac{\sqrt{\lambda}\, v^3 \alpha}{125}\sum_{i,j,k,l}\Bigl(-3G_4(i,j,k,l)+2G_{3}^{\rm sym}(i,j,k)-4 G_4(i,i,i,j)+G_4(i,i,i,i)\Bigr) \nonumber \\
=    \frac{\sqrt{\lambda}\, v^3 \alpha}{125}\Bigl(-72 \sum\limits_{i>j>k>l}G_4(i,j,k,l)+24 \sum\limits_{i>j>k}G_3^{\rm sym}(i,j,k)\nonumber \\ 
+42 \sum\limits_{i>j}G_4(i,i,j,j)-84\sum\limits_{i>j}G_4(i,i,i,j)-\frac{2080}{3}\Bigr)\,,
\end{gather}
where 
\begin{align}
G_3^{\rm sym}(i,j,k)   \equiv &\  G_4(i,i,j,k)+G_4(i,j,j,k)+G_4(i,j,k,k)
\nonumber \\  = & -2 \biggl(1 +\frac{R_i-R_j}{\tanh_{i-j}}+\frac{R_i-R_k}{\tanh_{i-k}}+\frac{R_j-R_k}{\tanh_{j-k}}\biggr)\,, \\
G_4(i,i,j,j) = &-4 \Bigl(1+\frac{1}{\sinh_{i-j}^2}\Bigl(1-\frac{R_i-R_j}{\tanh_{i-j}}\Bigr)\Bigr)\,, \\
G_4(i,i,i,j) = &-2 \frac{R_i-R_j}{\tanh_{i-j}}+\frac{2}{\sinh_{i-j}^2}\Bigl(1-\frac{R_i-R_j}{\tanh_{i-j}}\Bigr)\,, \\
G_4(i,i,i,i) = &-\frac{8}{3}\,.
\end{align}
This can be simplified to
\begin{align}
\lambda \alpha \int\limits_{-\infty}^{\infty}\diff y\,  \Tr\bigl(\hat T_{\rm wall}^4\bigr) = &
 \frac{\sqrt{\lambda}\, v^3 \alpha}{125}\biggl(-\frac{8560}{3}+64 \sum\limits_{i>j}\frac{R_i-R_j}{\tanh\bigl(R_i-R_j\bigr)}\nonumber \\ & -\sum\limits_{i>j}\frac{376}{\sinh^2\bigl(R_i-R_j\bigr)}\Bigl(1-\frac{R_i-R_j}{\tanh\bigl(R_i-R_j\bigr)}\Bigr)-72\sum\limits_{i>j>k>l}G_4\bigl(i,j,k,l\bigr)\biggr)\,.
\end{align}
Combining the above, we get the final result
\begin{align}
V_{\rm eff} =&  \frac{v^3\sqrt{\lambda}}{125}\biggl(
1000\mu^2 -\frac{8650}{3}\alpha+ \bigl(64 \alpha -100 \mu^2\bigr)\sum\limits_{i>j}\frac{R_i-R_j}{\tanh\bigl(R_i-R_j\bigr)} -\sum\limits_{i>j}\frac{376\alpha }{\sinh^2\bigl(R_i-R_j\bigr)}\nonumber \\ & \times\Bigl(1-\frac{R_i-R_j}{\tanh\bigl(R_i-R_j\bigr)}\Bigr)-72\alpha \sum\limits_{i>j>k>l}G_4\bigl(i,j,k,l\bigr)\biggr)\,.
\end{align}

\section{Proof of decompositon (\ref{eq:deco-A}) and (\ref{eq:decom-X})}\label{app:B}
\renewcommand{\theequation}{\Alph{section}.\arabic{equation}}

Here we give a proof that the extra-dimensional gauge field $A_{\alpha y}$ can be decomposed into
 $D_\alpha C_\alpha$ and the divergence-free component as in Eq.~(\ref{eq:deco-A}).
Firstly, we define a projection operator as 
\be
 P_\alpha=D_\alpha (\Delta_\alpha)^{-1}D_\alpha^\dagger\,.
\ee
This is a projection operator since it satisfies $P_\alpha^2=P_\alpha$.
Therefore we can always decompose $A_{\alpha y}=P_\alpha A_{\alpha y}+(1-P_\alpha)A_{\alpha y}$.
The first term is actually $D_\alpha B_\alpha$ in (\ref{eq:deco-A}) if we take 
 $B_\alpha = \Delta_\alpha^{-1}D_\alpha^\dagger A_{\alpha y}$.
However, $P$ is well defined only when $\Delta_\alpha^{-1}$ is well-defined.
A dangerous case occurs if $\Delta_{\alpha}^{-1}$ acts on its zero mode $\Delta_\alpha B^{(0)}=0~(D_\alpha B^{(0)})=0$
 as is given in (\ref{eq:mode-B}).
This is not the case, however, since $\Delta_\alpha^{-1}$ acts right after $D_\alpha^\dagger$.

Let $E$ be an arbitrary scalar and let us take an inner product
\be
 (B_0, D_\alpha^\dagger E)=(D_\alpha B_0, E)=0\,,
\ee
where we have defined the inner product by $(A,B)=\int dy A^*B$. 
Hence we conclude that $D^\dagger_\alpha E$ does not include $B_0$, and $\Delta_\alpha^{-1}D_\alpha^\dagger$
 is always well-defined, so is $P_\alpha$.
 
Let us next treat the second term $(1-P_\alpha)A_{\alpha y}$. 
This is denoted as $C_\alpha$ in (\ref{eq:deco-A}).
Indeed $C_\alpha$ satisfy the divergence-free condition (\ref{eq:div}):
\be
 D_\alpha^\dagger C_\alpha=D_\alpha^\dagger (1-P_\alpha)A_{\alpha y}
  =(D_\alpha^\dagger - D_\alpha^\dagger D_\alpha \Delta_\alpha^{-1}D_\alpha^\dagger)A_{\alpha y}=0\,.
\ee
This is a proof of the decomposition (\ref{eq:deco-A}).

We can repeat the same procedure for the decomposition of $\boldsymbol{X}$.
Let us first introduce the following projection operator
\be
P=\boldsymbol{D}\Delta^{-1}\boldsymbol{D}^\dagger\,.
\ee
Note that $\Delta^{-1}$ is well-defined because $\Delta=D_X^\dagger D_X+{\cal M}^2$. Then a vector
 $\boldsymbol{X}$ can be always decomposed into $\boldsymbol{X}=P\boldsymbol{X}+(1-P)\boldsymbol{X}$.
The first term is written as $\boldsymbol{D}Y$ if we defined $Y=\Delta^{-1}\boldsymbol{D}^\dagger X$.
The second term is clearly divergence-free since
\be
 \boldsymbol{D}^\dagger (1-P)\boldsymbol{X}=(\boldsymbol{D}^\dagger - \boldsymbol{D}^\dagger \boldsymbol{D}\Delta^{-1}\boldsymbol{D}^\dagger)\boldsymbol{X}=0\,.
\ee
Thus the second term is identified with $\boldsymbol{Z}$ in (\ref{eq:decom-X}).
This is proof of decomposition (\ref{eq:decom-X}).

Lastly we show that $\boldsymbol{Z}=(1-P)\boldsymbol{X}$ does not include the zero mode.
We decompose it as
\be
 (1-P)\boldsymbol{X}=\sum_{n\ge 0}\chi_n \boldsymbol{C}_n\,, \label{eq:X-expand}
\ee
where $\boldsymbol{C}_n$ stands for the eigenvector of $\tilde{\Delta}$ as
\be
\tilde{\Delta}\boldsymbol{C}_n=\tilde{M}_n^2\boldsymbol{C}_n\,.
\ee
We assign $n=0$ for the lowest eigenvalue of the non-negative definite operator $\tilde{\Delta}$.
Note that the lowest eigenstate is the kernel of $\boldsymbol{D}_R$, and it can be
 expressed as
\be
\boldsymbol{C}_0 \propto \boldsymbol{D}\Lambda\,,
\ee
with $\Lambda$ being an arbitrary function. 
Now let us verify if $(1-P)\boldsymbol{X}$ is orthonormal to $\boldsymbol{C}_0$ by the following
 calculation
\be
 ((1-P)\boldsymbol{X},\boldsymbol{C}_0) &\propto &((1-P)\boldsymbol{X}, \boldsymbol{D}\Lambda)\nonumber \\
 &=&(\boldsymbol{D}^\dagger (1-P)\boldsymbol{X},\Lambda)=0\,.
\ee
Hence the expansion (\ref{eq:X-expand}) does not include the lowest eigenstate $n=0$, namely $\boldsymbol{Z}$
 does not have the zero mode.
 
 \section{One loop RGE of SM}
 \label{app:C}
 The SM gauge couplings $\alpha_i={g_i^2 \over 4\pi}$ at one loop levels are given by
\begin{eqnarray}
 \alpha_i(\mu)^{-1}={1 \over \alpha_i(M_Z)}-{b_i \over 2\pi}\ln{\mu \over M_Z}\,,
\end{eqnarray}
where $i=1, 2, 3$ correspond to the gauge goups $U(1), SU(2), SU(3)$, respectively. 
Here $b_1={41 \over 10}, b_2=-{19 \over 6}, b_3=-7$ and $\alpha_1(M_Z)={5\alpha_{\rm EM} \over 3\cos^2\theta_W}$,
 $\alpha_2(M_Z)={\alpha_{\rm EM} \over \sin^2\theta_W}$, $\alpha_3(M_Z)=0.1176$ where 
 $\alpha_{\rm EM}={1 \over 127.909}$ and $\sin^2\theta_W=0.23119$.
 
 The RGEs for Yukawa couplings are given by
 \begin{eqnarray}
  {d \over dt}H_a={1 \over 16\pi^2}\beta_a^{(1)}H_a+H_a{1 \over 16\pi}\beta_a^{(1)\dagger}\,,
 \end{eqnarray}
where $a=u, d, e$ and $t=\ln(\mu/M_Z)$.
$\beta_a^{(1)}$ is give as
\begin{eqnarray}
 \beta_a^{(1)}=c_a^{(1)}{\bf 1}+\sum_b a_a^b H_b\,,
\end{eqnarray}
where $c_a^{(1)}=T_a-G_a$ with
\begin{eqnarray}
 G_u&=&{17 \over 20}g_1^2+{9 \over 4}g_2^2+8g_3^2\,, \\
 G_d&=&{1 \over 4}g_1^2+{9 \over 4}g_2^2+8g_3^2\,, \\
 G_e&=&{9 \over 4}g_1^2+{9 \over 4}g_2^2\,, \\
 T_u&=&T_d=T_e=3{\rm Tr}(H_u+H_d)+{\rm Tr}H_e\,, \\
 a_u^u&=&a_d^d={3 \over 2}\,,\\
 a_u^d&=&a_d^u=-{3 \over 2}\,, \\
 a_e^e&=&{3 \over 2}\,.
\end{eqnarray}
%
%
%
\noindent {\bf Acknowledgements} \\
\noindent The authors (M.A. and M. E.) wish to acknowledge the support of the KA171/Erasmus+ programme, which made it possible to 
 stay at the Silesian University in Opava, where this work was brought to completion.
This work is supported in part by JSPS Grant-in-Aid for Scientific Research KAKENHI 
 Grant No. JP21K03565 and JP25K07275 (M. A.).
This work has been also supported by the grant no. SGS/24/2024 Astrophysical processes in strong gravitational and electromagnetic fields of compact object (F. B.).


\end{document}